\documentclass[prodmode,acmtecs]{acmsmall} 

\usepackage[ruled]{algorithm2e}
\usepackage{longtable}
\usepackage{caption}
\newcommand{\subparagraph}{}
\usepackage[compact]{titlesec}
\titlespacing{\section}{0pt}{*0}{*0}
\titlespacing{\subsection}{0pt}{*0}{*0}
\titlespacing{\subsubsection}{0pt}{*0}{*0}

\SetAlFnt{\small}
\SetAlCapFnt{\small}
\SetAlCapNameFnt{\small}
\SetAlCapHSkip{0pt}
\IncMargin{-\parindent}

\usepackage{color}
\usepackage[x11names]{xcolor}

\acmVolume{1}
\acmNumber{1}
\acmArticle{1}
\acmYear{2016}
\acmMonth{4}

\doi{0000001.0000001}

\issn{1234-56789}

\begin{document}

\markboth{Yadav et al.}{ Mining Electronic Health Records (EHR): A Survey}

\title{ Mining Electronic Health Records (EHR): A Survey}
\author{Pranjul Yadav
\affil{ University of Minnesota - Twin Cities}
Michael Steinbach
\affil{ University of Minnesota - Twin Cities}
Vipin Kumar
\affil{University of Minnesota - Twin Cities}
Gyorgy Simon
\affil{University of Minnesota - Twin Cities}
}

\begin{abstract}
The continuously increasing cost of the US healthcare system has received significant attention. Central to the ideas aimed at curbing this trend is the use of technology, in the form of the mandate to implement electronic health records (EHRs). EHRs consist of patient information such as demographics, medications, laboratory test results, diagnosis codes and procedures. Mining EHRs could lead to improvement in patient health management as EHRs contain detailed information related to disease prognosis for large patient populations. In this manuscript, we provide a structured and comprehensive overview of data mining techniques for modeling EHR data. We first provide a detailed understanding of the major application areas to which EHR mining has been applied and then discuss the nature of EHR data and its accompanying challenges. Next, we describe major approaches used for EHR mining, the metrics associated with EHRs, and the various study designs. With this foundation, we then provide a systematic and methodological organization of existing data mining techniques used to model EHRs and discuss ideas for future research. \\
\end{abstract}

%
%

%
%



\acmformat{Pranjul Yadav, Michael Steinbach, Vipin Kumar, Gyorgy Simon,2016. Mining Electronic Health Record (EHRs) : A Survey}

\begin{bottomstuff}
\end{bottomstuff}

\maketitle

\section{{Introduction}}
Numerous recent studies have found that the health care system in the United States is the most expensive
in the world, yet it is trailing behind most advanced economies in quality \cite{schuster1998good}.  The cost of health care is
steadily rising both in absolute terms and also as a percentage of the GDP, soon reaching unsustainable levels \cite{levit2003trends}.
To curb this trend, the US government has put several initiatives in place, aiming to
simultaneously improve care quality and decrease its cost.
Many believe that advanced analytics holds the key to achieving these opposing goals. 

The earliest of these initiatives was the mandate for health care providers to
implement electronic health record (EHR) systems. While the primary goal of the EHR is to
documenting patients' care for reimbursement, a \emph{secondary use} of the accumulated 
data is often to serve as a research platform.  This EHR data coupled with other health-related data
provides the platform on which advanced analytics can be built.

The motivation for advanced analytics comes from another government mandate that transforms 
the current fee-for-service payment model to a new model based on population health management.
Under the new model, primary care providers are responsible for managing entire patient populations with their payments tied to care quality. Since care providers are no longer paid for services rendered, but rather for the outcomes of the service, they are incentivized to increase their
efficiency through implementing better care practices, thus creating an opportunity for analytics. 

With ample data in EHR and a motivation for analytics, we need an outlet for the findings. 
The learning health care system, a concept embraced by the Institute of
Medicine, is an adaptive system, in which ``best practices seamlessly embedded in the delivery process and new knowledge captured as an integral by-product of the delivery experience".
In this context, advanced analytics, clinical data mining included, will without doubt play an
important role in extracting new knowledge, quantifying the effects of changes in care delivery
and possibly forming best-practice guidelines.

The learning health care system initiative is accompanied by other symbiotic initiatives that
are likely to influence the medicine of the future.  Precision medicine is one of them. It is a model,
where treatment is tailored to each individual patient. While precision medicine has a large
genomic component, finding increasingly specific treatments from EHR data is another
avenue, one that naturally fits into today's practice framework.

The predominant medical practice framework today is evidence-based medicine\cite{sackett2000evidence}.
At the heart of evidence-based practice lies a large knowledge base of best practice recommendations
that have been put forth by committees of well-established care providers and (ideally) validated through randomized clinical trials. These best practice recommendations, if followed, can enable providers to increase their efficiency and reduce waste. 

The traditional vehicle for evidence creation is the randomized clinical trial \cite{matthews2006introduction}.
Randomized clinical trials are considered the gold standard for evidence creation,  but the body of evidence they create 
is woefully incomplete.  Trials are very expensive to conduct, thus they are relegated to a \emph{confirmatory}
role; in particular they are impractical for exploration.

Clinical data mining on top of EHR data offers a solution which makes it  complementary 
to clinical trials.  With EHRs tracking entire populations over extended time periods, EHR data have large
sample sizes and potentially long follow-up times possibly allowing us to test a large number of 
exploratory hypotheses.
Through flexible modeling, EHR data can provide useful information to facilitate clinical decision support, to analyze condition-specific clinical process outcomes and to improve team-based population care beyond the tradional clinical encounters.

With all three ingredients of success for data mining---data, motivation and outlets for results---having come together, we believe that an explosive growth in the adoption of clinical data mining is imminent.
Unfortunately, EHR data poses unique challenges that data mining has previously not faced.
Before reviewing existing work on clinical data mining involving EHR data, in this survey, we summarize background knowledge on study design, the characteristics and challenges of EHR data, and techniques developed by other disciplines to overcome
these challenges. We hope that our survey will prove to be a useful tool that helps turning the 
challenges that EHR data poses into a wave of innovation in data mining that will unlock the 
immense potential of clinical data mining. 

\vspace{1mm}\noindent\textbf{Overview:} Our survey is structured as a reflection of the steps carried out in clinical research.
Our first step is to frame a clinical problem that we wish to study.
In Section 2, we give the reader a flavor of the clinical questions by providing descriptions of 
several applications starting from the simplest clinical questions and progressing towards more 
complex analytics tasks.

Once the clinical question has been defined, data modeling starts. 
In Section 3, we provide a detailed description of EHR data, its elements and characteristics.
As we discussed earlier, the primary function of EHR data was not to serve as a research 
platforms thus it poses numerous challenges which we describe in  Section \ref{sec:DataChallenges}. 

As the next step in the analysis process, we select a metric of interest.
EHR data is multidimensional, multimodal irregular time series data with many potential outcomes. 
A rich array of metrics                                                                                                                                                                                                                                                                                                                                                                                                                                                                                                                                                                                                                                                                                                                                                                                                                                                                                                                                                                                                                                                                                                                                                                                                                                                                                                                                                                                                                                                                                                                                                                                                                                                                                                                                                                                                                                                                                                                                                                                                                                                                                                                                                    can be designed and measured on such data, but only a handful are well defined, understood and accepted in the health sciences research community. 
In Section 5, we describe the most commonly used metrics, those that the community tracks
and uses to inform policy decisions. 

EHR data is a collection of data tables, a format that data mining algorithms cannot directly operate on.
These tables need to be summarized into a data matrix in
a manner that allows us to answer our clinical question or compute our metric of interest. 
This process is known as study design and in Section \ref{sec:StudyDesign}, we will introduce the study designs that are most suitable for mining EHR data.

At this point in the analysis, we have a data matrix that we could run our data mining algorithms on,
but doing so has the potential to produce misleading results. EHR data still has some challenges that
could seriously affect or invalidate our analysis. 
Examples of such challenges include situations when a patient drops out before the study concludes (referred to as censoring); patient populations whose members are in widely different states of health and thus outcomes are not comparable across these patients; and a highly variable number of per-patient observations. 
In Section \ref{sec:Approaches}, 
we introduce higher level frameworks that have been developed to overcome such challenges.

Data mining algorithms can operate within these frameworks or can be adapted to fit into these
frameworks. In section \ref{sec:survMeth},
we provide a comprehensive overview of how data mining methods  have been applied to mine {EHR} data and how the data mining methods have contributed to the clinical research carried out in various applications areas, that we introduced in Section 2. 

In Section 9,  we present the current state of clinical data mining and identify areas that are well covered
by existing techniques and areas that require the development of new techniques and finally, while Section 10 concludes with the discussion of future of EHR data mining.

Although health care data mining is still in its infancy, in this survey we have covered several hundred
scientific articles in this area.  We describe the required background knowledge and lay out the principles that allow us to organize this vast body of research. Due to page limitations, we have focused the main article on the sequence of steps researchers need to take to complete an EHR-based data mining analysis citing only a small number of representative examples. Accompanying supplemental material contains a more comprehensive listing of papers descibing application of data mining to EHR data.

\vspace{1mm}\noindent\textbf {Scope:} The primary purpose of this survey is to provide the necessary background 
and an overview of the existing literature of mining structured EHR data towards answering clinical questions. 
Areas, such as bioinformatics, computational biology and translational bioinformatics, which are indispensable
to the future of medicine, but utilize their own specific techniques and deserve surveys of their own right; 
we decided to exclude them from the scope of this survey. We also excluded medical image analysis and techniques
related to mining semi-structured or unstructured data through Natural Language Processing (NLP) and information retrieval because the technical challenges they pose are quite different from the
challenges we face in mining structured EHR data.
\\

 \section{{Application Areas}}\label{sec:Introduction}
 
An appropriate place to start our discussion of clinical data mining is to describe some clinical questions
that data mining can help answer.
The fundamental question of medicine is to decide on the treatment that is most suitable and effective for a particular patient and data mining has the potential to help address this question in numerous
ways.

In this section, we present a sequence of increasingly complex clinical applications, 
starting with the simplest epidemiological questions,
continuing with common clinical research tasks such as automatic identification of study cohorts,
risk prediction and risk factor discovery, all the way to complex applications such as
discovering knowledge related to best clinical practices from data. 

 \subsection{Understanding the Natural History of Disease}
The most basic epidemiological inquiries are not data mining problems per se, they are concerned with
questions like: How many patients are diagnosed with a condition of interest? What is the incidence rate and symptoms of a disease?  What is the \emph{natural history of progression of the disease}, or in other words, what are the \emph{symptoms and implications} of any clinical condition of interest? 

Studying the sequences of diseases clearly enters the realm of data mining and it has been applied to studying the progression of a patient\rq s medical state over time, which is also known as the patient\rq s \emph{medical trajectory}. Examples of such trajectories are the progression of the patient from a healthy state through conditions like hyperlipidemia (high cholesterol), hypertension (high blood pressure), diabetes towards diabetes associated complications (e.g. amputation,  severe paralysis or death). Often, multiple trajectories lead to the same outcome. For example, consider an outcome such as mortality. In this case, a patient might die due to kidney complications, cardio-vascular complications or peripheral complications. Even though the outcome (mortality in this case) is the same, disease progression paths leading to the outcome might be different. Research studies have demonstrated that different trajectories can have significantly different associated risks for the same outcome \cite{oh2016,yadav0231}. Studying such varying trajectories can lead to the development of tailored treatments, discovery of biomarkers or the development of novel risk estimation indices. 

The number of patients suffering from two or more chronic diseases, a condition referred to as \textit{multimorbidity}, has increased substantially in the last ten years and multimorbidity is the norm with elderly patients \cite{mercer2009}. Diseases co-occur due to coincidence, causal relationships and common underlying risk factors including frailty. The coexistence of multiple chronic diseases increases the risk of mortality, physical function, decreases quality of life and is also associated with longer hospital stays, increased risk of postoperative complications and a higher overall healthcare utilization \cite{wolff2002}. 
When diseases co-occur, it is desirable to treat them simultaneously. \textit{Comorbidity analysis} is the process of exploring and analyzing relationships among diseases and it is the key to understanding multimorbidity and tailoring treatment accordingly. An example of multimorbidity is type 2 diabetes mellitus (T2DM) which is often accompanied by hypertension, hyperlipidemia and impaired fasting glucose (IFG); these latter three conditions are \textit{comorbid} to T2DM. Further, analyzing the comorbidities and discovering the relationships among them can lead to the modification of existing comorbidity scores (such as Charlson index) or to the development of novel ones.
 
\subsection{Cohort Identification}

Once we understand these fundamentals,  we can try answering more advanced questions. To this end,  we may need to assemble a cohort (group) of patients, some of whom are extremely likely to have the
disease of interest (\emph{cases}) and others who most likely do not (\emph{controls}). This can be achieved through phenotyping algorithms, either hand-crafted or machine learned. Phenotyping algorithms characterize the disease in terms of patient characteristics observable from the EHR data and classify patients as likely having the disease, likely not having the disease, or disease status is uncertain \cite{Kirby:2016}. 

Traditionally, cohort identification was carried out through chart reviews, where nurse abstractors have painstakingly reviewed patients' medical records to identify whether each patient satisfies the criteria for
inclusion into the cohort. However the scale enabled by EHRs renders manual chart review impractical. Instead, electronic phenotyping algorithms are applied, with manual chart review relegated to spot-checking. Cohort identification has been widely used in various clinical research studies and biomedical applications, forming the platform on which future studies can be carried out in areas such as pharmacovigilance,  predicting complications, and quantifying the effect of interventions. 

A phenotype is defined as a biochemical or physical trait of an organism, such as a disease, physical characteristic, or blood type. 
Examples of phenotypes in EHRs are clinical conditions, characteristics or sets of clinical features that can be determined solely from the EHR data. Such techniques are useful for identifying patients or populations with a given clinical characteristic from EHRs using data that are routinely collected and stored in disease registries or claims data. Phenotyping queries used for cohort identification can be used at different sites in a similar fashion in order to ensure that populations identified across  healthcare organizations have similar clinical state. Phenotypic definitions can also be used for direct identification of cohorts based on  risk factors, clinical or medical characteristics, and complications, thereby allowing clinicians to improve the overall healthcare of a patient. For a detailed review of phenotyping, the reader is referred to Shivade et al. \cite{shivade2014review}.

 \subsection{Risk Prediction/Biomarker Discovery}
 
With a cohort in hand, we can build risk models, opening up a wide range of opportunities for data mining. An example of a successful risk model is the Framingham heart score which estimates
patients' risk of cardio-vascular mortality. In recent years, age-adjusted cardio-vascular
deaths have reduced by half in developed countries. Much of this success is attributed
to the Framingham Heart Study, which helped identify the key risk factors of
cardio-vascular mortality \cite{bitton2010framingham}.
This success can potentially be replicated in other areas of medicine.

 Such models can predict the risk of developing a disease, e.g. estimating the probability of developing a condition of interest in 5 years  (\emph{risk prediction}). Such analysis is often performed to identify high risk individuals, and facilitate the design of their treatment plans \cite{ng2014paramo}. Interventions prescribed by risk analyses can lead to improvement in a patient's health, thereby preventing the patient from progressing to advanced complications. 

In some cases, predicting the patient's risk of progression is secondary to understanding the underlying risk factors. Risk models can provide information about the importance of  risk factors. 
Risk prediction also provides the opportunity to identify significant indicators of a biological state or condition. In simple terms, a biomarker is defined as a set of measurable quantities that can serve as an indicator of a patient's health. Biomarkers offer a succinct summary of the patient\rq s state with respect to a medical condition. Rather than having to analyze the thousands of variables present in an EHR, it can be sufficient to focus on relatively few biomarkers to paint a reasonably accurate picture of the patient's overall health. Over the years, biomarkers have found numerous applications. For example, abnormal hemoglobin A1C (measure of blood sugar) is a biomarker for Type-2 Diabetes Mellitus (T2DM) and high cholesterol is a biomarker for being at risk of cardio-vascular complications. Similarly, there are certain biomarkers, which are common across many diseases. For example, age is by far the most common biomarker. It indicates that as a person ages, his or her risk to acquire certain diseases (e.g T2DM, cardio-vascular complications and kidney complications) increases significantly. EHRs provide a platform to identify, analyze and explore biomarkers for multiple diseases. 

\subsection{Quantifying the effect of Intervention}
Interventions are often drug therapies or surgeries,  but can also include recommendations for life style changes and/or patient education.  Choosing the optimal
treatment for a patient requires us to be able to estimate the effect of the possible interventions. Specialized data mining methods such as uplift modeling or statistical techniques in combination with causal analysis can be used to \emph{quantify the effects of interventions}. 

The longitudinal aspect of EHRs provides an opportunity to analyze the effects of intervention for longer period of time across larger cohorts. It also provides clinicians with a platform to analyze whether the interventions have any accompanying adverse effects. Moreover, EHRs provide a platform to analyze whether interventions vary across cohorts based on demographics attributes such as gender, age, ethnic make-up, socio-economic status, etc.

\subsection{Constructing Evidence Based Guidelines }

Once the effect of a treatment has been proven,  this knowledge can be
codified into and disseminated as clinical practice guidelines. \emph{Evidence-based 
clinical practice guidelines} are considered the cornerstone of modern medicine,  and they provide guidance on the optimal treatment under a particular set of conditions
based on epidemiological evidence. 

Clinical guidelines can be defined as standardized specifications and procedures usually used to improve or take care of a patient in specific clinical circumstances \cite{field1992guidelines}. Evidence based guidelines try to guide decision making by identifying best clinical practices, that are meant to improve the quality of patient care \cite{barretto2003linking}. They help clinicians make sound decisions by presenting up to date information about best practices for treating patients in a particular medical state including expected outcomes and recommended follow up interval. For example, the American Diabetes Association (ADA) guidelines for diabetes consist of recommendations for diagnosing the condition (e.g. a patient is considered diabetic if his hemoglobin A1c is greater than 6.5), controlling the disease (patient is under control if his A1c $<$ 6.5 and his systolic blood pressure $<$ 140 mmHg) and prescribing interventions (lifestyle modification and therapeutic interventions, as well). These guidelines are often regarded as the cornerstone of modern healthcare management.

  \subsection{Adverse Event Detection}
This term describes the detrimental effect of patient medical state as a result of medical care. Examples include 
infection acquired during the treatment of a different condition, such as surgical site infection.
According to a 2010 report by the Inspector General of the Department of Health and Human Service, 
13.5\% of hospitalized Medicare beneficiaries have experienced adverse events, costing Medicare an estimated \$340 million in Oct. 2008 alone. 
An estimated 1.5\% of beneficiaries, which corresponds to 15,000 patients per month, experienced an adverse event that lead to their deaths \cite{levinson2010adverse}.
Detecting and learning to prevent adverse events is clearly a major opportunity for advanced analytics, as 44\% of these adverse event were deemed avoidable.

Another related opportunity is the identification of potentially preventable events.
Potentially preventable events are patient encounters (emergency and urgent care visits, or
hospitalization) that could have been avoided by appropriate outpatient treatment and adequate 
compliance with those treatments.
The Agency for Healthcare Research and Quality (AHRQ) reports that potentially preventable
events are decreasing, but still nearly 4 million hospitalizations in the US were avoidable in 2010, if the patient had received proper care \cite{torio2006trends}.
Costs related to these preventable hospitalizations totaled \$31.9 billion in 2010.

Adverse events can often be linked to drugs. 
Adverse drug events account for 1 in 3 hospital adverse events and
in the outpatient setting, adverse drug events cause 3.5 million physician visits, 1 million
emergency room visits and 125,00 hospitalizations each year \cite{torio2006trends}.
Although drugs are tested for any potential adverse effects before they are released for widespread use, often test cohorts are small with short observation periods. Several agencies conduct research on detecting adverse drug reactions: U.S. Food and Drug Administration with its adverse event reporting system, the European Medicines Agency, and the World Health Organization, which maintains an international adverse reaction database. Despite these efforts, all these agencies suffer from underreporting and biased analyses of adverse drug reactions. EHRs provide a new platform to improve and complement drug safety surveillance strategies.

\section{Nature of EHR Data}\label{sec:EHRData}
One motivation behind the federal mandate for EHRs was to document patients' state of health over time and the therapeutic interventions to which these patients were subjected. EHRs store this information in structured (databases), semi-structured (flow sheets) and unstructured formats (clinical notes). The format of the information greatly affects the ease of access and quality of the data, and thus has substantial impact on the downstream data mining.

\subsection{Structured Data}
From the viewpoint of healthcare analytics, retrieving structured data is the most straightforward. Structured data is stored in database tables with a fixed schema designed by the EHR vendor. The most commonly used information, such as demographic information (e.g. birth date, race, ethnicity), encounters (e.g. admission and discharge data), diagnosis codes (historic and current), procedure codes, laboratory results, medications, allergies, social information (e.g. tobacco usage) and some vital signs (blood pressure, pulse, weight, height) are all stored in structured tables. This kind of information is common across providers and not specific to any clinical specialty. Thus the use and format of this information is well handled by the EHR vendors. This allows such information to be stored in structured data tables with apriori defined layouts (schema). Fixed schemas enable high performance (rapid access to data) and standardization: the schemas for these tables are very similar if not identical across installations by the same EHR vendor, requiring very little (if any) site-specific knowledge from users. This quasi-standardization of fields also greatly helps information retrieval for analytic purposes.

Storing all information in EHRs as structured elements, however, is impractical: it would require anticipation of all possible data elements (e.g. metrics whose usefulness we do not yet appreciate) and would result in a level of complexity that would render the EHR system unusable. However, there is a need for storing information that does not readily fit into the admittedly rigid schema of the structured tables. For example, clinicians often write notes about patient\rq s symptoms based on their previous experiences, which is hard to standardize a priori.

\subsection{Unstructured Data}
Among the three formats, clinical notes (unstructured data) offer maximal flexibility. Clinical notes mostly store narrative data (free text). Many types of clinical notes are in existence, and the information that resides in them depends on the type of note (e.g. radiology report, surgical note, discharge notes). These clinical notes can contain information regarding a patient's medical history (diseases as well as interventions), familial history of diseases, environmental exposures and lifestyle data all reside in clinical notes. Natural language processing (NLP) tools and techniques have been widely used to extract knowledge from EHR data. 

Clinical notes such as admission, treatment and discharge summaries store valuable medical information about the patient, but these clinical notes are very subjective to the doctor or the nurse writing them, and lack a common structure or framework. These clinical notes also have grammatical errors, short phrases, abbreviations, local dialects and misspelled words. Considerable data processing needs to be conducted on these clinical notes such as spelling correction, word sense disambiguation, contextual feature detection, extraction of diagnosis codes from clinical text, and adverse events surveillance. This makes deriving information about patient characteristics from clinical notes a computationally challenging task that requires the most sophisticated NLP tools and techniques. For a detailed review on application of NLP to clinical decision support, the reader is referred to \cite{demner2009can}

\subsection{Flowsheets}
In between the two extremes (structured tables and unstructured clinical notes) lies the (semi-structured) flow sheet format. This format is most reminiscent of resource description files (RDF), consisting of name, value and time stamp triplets. Typically, the ``name'' field stores the name of the measure and the ``value'' field contains the actual measurements: e.g. the name is ``arterial blood pressure" and the value is 145 Hgmm. This format is more flexible than the structured tables, since the user can define new metric through the name field; the set of metrics is not restricted to those anticipated by the EHR vendor. Flow sheets are similar to structured data in the sense that the value field is either a quantitative measure (e.g. blood pressure) or typically a restricted set of values. For instance, the American Society of Anesthesiologists (ASA) physical status takes values of  ``mild systemic disease",``healthy", ``severe systemic disease",``moribund" or ``severe life-threatening systemic disease". 

Flow sheets offer expandability to EHR systems and thus have found numerous uses, becoming the only or most convenient data repository for many applications. Possibly the most important use for flow sheets is that they provide detailed information about specialty care. For example, information related to a patient's asthma care plans can be stored in flow sheets or they may store various diabetes-related non-standard (or not-yet-standard) metrics for a diabetes clinic. In addition, they may provide additional details regarding how a particular measure was obtained (blood pressure taken while the patient was lying flat) and can also be used to store automated sensor data (e.g. pulse and blood oxygen levels every few minutes in an intensive care unit). Further, flow sheets can be used to pull together related measurements such as quality indicators.\\

\section{Data-Related Challenges}\label{sec:DataChallenges}
EHR data as a research platform poses numerous challenges. Many of those challenges are not specific to EHR data and are also frequently encountered in other domains: noise, high dimensionality, sparseness, non-linear relationships among data elements, complicated dependencies between variables. Less frequent, but still not uncommon elsewhere, are issues related to data integration across multiple sites (medical providers) and / or multiple types of data sets (e.g. clinical vs. claim data). As in other domains, it is also important to incorporate domain knowledge, including knowledge about the relationships. For example, the blood pressure of a patient on medication for hypertension needs to be interpreted in that context. 

In this section, we describe a number of challenges, common in health care but less studied in other areas, that put the external validity of the analysis at risk. Many of these relate to issues stemming from data missing for various reasons. 

In the following subsections, we will discuss these various issues in greater detail.

\subsection{ Censored Data}
By censored data, we refer to the situation where  a patient's state is only observable during
a certain period of time; or conversely, when potentially interesting events fall outside the
observation period and are hence unobservable. In case of left censored data, patients experience events of interest prior to the start of the study;
in case of right censored data, potentially interesting events are unobservable because they happened to the
patient after he dropped out of the study or after the conclusion of the study. In case of interval censored data, information is only available of the data being within a certain limit. Studies can be either left, right or interval censored. Censoring can lead to loss of crucial information about the patient\rq s health. For example, for the right censored patient, there is neither an easy way to determine whether the patient is alive or dead nor to measure the efficacy of the treatment the patient was undergoing. 

\subsection{Fragmentation}
Fragmentation is a lack of data sharing across providers. Fragmentation typically occurs when patients visit multiple healthcare providers seeking specialty care, expert advice or second opinions. In such scenarios, all healthcare provider involved only have partial information about the patient's medical history. Integrating data across multiple healthcare providers poses several challenges. Some of these are operational challenges,  such as communication between different EHR systems such as General Electric (GE) or Epic,  requiring a common language to transfer information into HL-7 (common protocol), which cannot capture all nuances.  Even when multiple sites use the same EHR format, their treatment policies may differ, flowsheets may differ and thus their definitions of nuanced concepts may differ. For example, fasting and random glucose measurements are not distinguished by lab codes and different sites can apply different methods to distinguish the two. Other challenges involve the (lack of) willingness of competing providers to share data.

\subsection{ Irregular Time Series Data}
Beside our inability to make observations before the study period starts or after it concludes, the most striking characteristic of the EHRs data is the irregularity of the patient visits. While recommended frequency of visits may exist, many patients do not actually follow these recommendations. For example, as per the American Diabetes Association (ADA) guidelines, A1C test must be performed at least two times a year for patients who are meeting treatment goals and have stable glycemic control.

Further, the frequency at which information is collected varies. While vitals are collected at every visit, certain laboratories tests are ordered annually, and other tests are performed only as needed. This difference in the frequency of collection of medical information leads to irregular longitudinal data. Analyzing regular time series is a well-studied problem in data mining, but application of these techniques to EHR-type irregular time series is very challenging.

\subsection{Other Sources of Missing Data}
Diagnosis codes might also be missing due to reimbursement rules. Different problems, comorbidities or complications have different reimbursement rates:  depending upon the complications, the same procedure may have different costs and thus result in increased or decreased reimbursements. Due to these financial constraints only some of the problems related to the primary cause of the visit, are used to generate billing codes (ICD codes are used to represent these problems in billing records). This leads to biases in ICD-9 codes, as the billing codes might not be a true representation of the actual medical state of the patient.

Diagnosis codes can also be missing due to changes in disease definitions and updates to the ICD-10 codes. For example, pre-diabetes did not have a corresponding ICD-10 code until 2000. The introduction of new and the periodic updates to existing ICD codes leads to further complications such as a lack of a clear mapping from the old revision to the new and subsequently, to inconsistent research findings.

Another largely unobservable source of missing information lies in patient conformance with prescriptions and other intervention, such as lifestyle change recommendations. The orders table in an EHR indicates that the physician prescribed a medication, but in most cases we do not know whether the patient actually took the medication.  This situation is referred to as Intent to Treat. In the case of the lifestyle change, we may not even have documentation that the patient followed this advice. 

A unique aspect of missing data in clinical analytics is that whether the data is missing or not can be predictive. When a physician orders a test, he usually suspects that the patient may suffer from the corresponding condition. Conversely, by not ordering certain tests, the clinician suggests that corresponding medical conditions are absent. For example, no bacterial panel being ordered likely indicates that the patient is not suffering from any infection.

\subsection{{Biases and Confounding Effects}}
Studies performed using EHRs often have biases and confounding effects \cite{moher1998does}. Biases might arise due to multiple reasons. For example, in a cohort study, there might be significant differences in baseline characteristics (age, gender, race, ethnicity) between the cases and the controls \cite{gruber1986clinical}. In such cases, any observed difference between the groups after a follow-up period might be due to the difference in baseline characteristics and not due to the exposure. Therefore in such cases, quantifying the real effect of exposure might be difficult. 

Such bias can be overcome by finding the right control group. One possible way is to randomly select subjects from a pool of patients such that the pool does not consist of patients diagnosed with the outcome \cite{wacholder1992selection}. In other approaches, controls can be drawn from neighborhood of the cases as such controls would be very similar in terms of socio-economic status and lifestyle choices \cite{vernick1984selection}. Similarly, when genetic factors are the main focus of study, controls could often be chosen from family and relatives as they share similar genetic make-up \cite{wacholder1992selection}. 

Confounding is another issue which might undermine the internal validity of any study \cite{abramson2001making}. Confounding arises when a variable (i.e. confounder) is associated with the exposure and influences the outcome, but the confounding variable is not connecting the exposure with the outcome \cite{ory1977association}. For example, studies have often reported a high degree of association between risk of myocardial infarction and oral contraceptives. However, this association was later found to be spurious because of the high proportion of tobacco users among users of birth control pills. Therefore tobacco consumption confounded the relation between myocardial infarction and oral contraceptives. \\

\section{Metrics}\label{sec:Metrics}
Quantifying the outcome is the primary interest in many research studies. The outcome is often quantified using various metrics such as incidence rate, prevalence, relative risk and odds ratio \cite{hennekens1987epidemiology}. Incidence rate \cite{last2001dictionary} indicates the number of new cases of disease in a population at risk over a predefined interval of time. Prevalence indicates the number of existing cases of disease in a population under observation. For example, in a given population of 100,000 persons, there were 980 patients who have the diagnosis of tuberculosis and there happen to be 10 patients freshly diagnosed with tuberculosis in the same year. In this scenario, the incidence rate of tuberculosis within a year would be 10/99020  whereas the prevalence rate would be 980/100000. \\
Relative risk  is defined as the frequency of outcome in the exposed group as compared to the frequency of outcome in the unexposed group. For example, consider a cohort of pre-diabetic patients. The cohort is divided into two groups (control and treatment) of 1000 patients each. The treatment group is prescribed statin and the control group is not. The cohort is then followed for 5 years. After 5 years, it was observed that 200 patients in the treatment group progressed to diabetes whereas 100 patients in the control group progressed to diabetes. From this information, the relative risk of diabetes is 2.0: patients within treatment group are twice as likely to progress to diabetes as controls. Relative risk is 1.0 when the frequency of outcome is same in both the groups. Relative risk greater than 1.0 indicates increased risk of outcome in the treatment group, while less than 1.0 indicates decreased risk (protective effect of exposure). \\
Odds ratio can be defined as the odds of exposure/outcome among the intervention group divided by the odds of the exposure/outcome among the non-inervention group. For the example above, the odds in the case group will be 0.20 whereas the odds in the control would be 0.10. The odds ratio would then be 2.0. In the following section, we describe various study designs and metrics which are widely used in respective study designs. \\

\section{{Study Design} }\label{sec:StudyDesign}

{EHR} data is a mere collection of database tables that need to be transformed into an analysis matrix that
is amenable to data mining in a manner that allows us to answer the question we set out to study.
Suppose we wish to construct a risk prediction model that can predict the 5-year risk of a particular
disease $D$ for any patient. To construct such a model, we would take a cross section of the patient
population in year $Y$, allowing us to have a representative sample from which this ``any patient'' may come
from. $Y$ is ideally at least 5 years before the last date in the {EHR}, so that we
have sufficient (5-year) follow-up for many patients. Next, we reconstruct all patients' state of health
in year $Y$, collecting their medical history before $Y$; and then follow them five years forward (until year
$Y+5$) and establish their outcomes, namely whether they developed $D$. The analysis matrix
would have patients as its rows and patient characteristics in year $Y$ as columns; and would have
an outcome for $D$, as well. Traditional predictive modeling techniques are directly applicable to such a matrix for carrying out various research related tasks. 

The way we transformed that {EHR} data follows a particular \emph{study design} 
that allows us to answer our question. Each study question can require a different study design.
In this section, we review some of the most commonly used study designs and the questions they allows us to answer. For additional details, the interested reader is referred to \cite{grimes2002overview}.

\begin{figure}
\centering
\includegraphics[width=5cm, height=5cm]{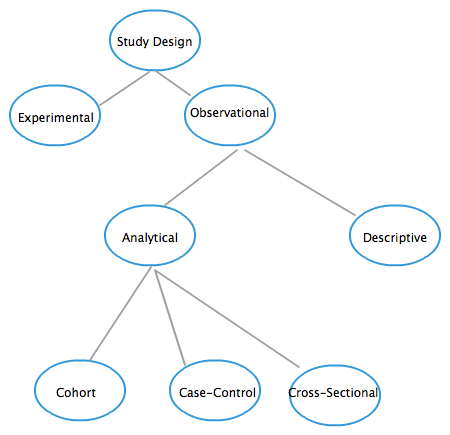}
\caption{Study Design Classification Hierarchy \label{overflow}}
\end{figure}

Study designs form a hierarchy which is depicted in Figure \ref{overflow}. Accordingly, at the highest level, study designs
can be primarily classified into two major groups i.e. experimental and observational.  

In an \textbf{experimental} study design,  the researcher intervenes to change the course of the disease and then observes the resultant outcome. Randomized Clinical Trials are examples of experimental study designs. A specific example would be a study where surgery patients with T2DM were randomized to receive supplemental insulin at bedtime for blood glucose (treatment) or no supplemental insulin (control). As intervention in EHRs is not possible,  we will not discuss these study designs any further. 

By \textbf{observational} \cite{funai2001distribution},  we refer to study designs where the researchers do not intervene. In such studies, the treatment alloted to each subject is beyond the control of the investigator. For example,  consider a study that investigates the effect of smoking (exposure) on lung capacity (outcome). A cohort of young men aged 18-25 are identified. Some subjects in this cohort smoke tobacco (exposed group) and some do not (unexposed/comparison group); the investigator has no influence on which subjects smoke and which do not. This cohort is then followed for a number of years to analyze the effect of smoking on lung capacity by comparing the exposed group with the unexposed group \cite{kelsey1996methods,rothman1986modern}. Observational studies can be further categorized as analytical (if there is a comparison group as in the example given above) or descriptive (no comparison group). 

\textbf{Analytical} studies are mostly used to test hypotheses about exposures and their effects on the outcome. They can also be used to identify risk and protective factors for diseases as well as causal associations between exposures and outcomes. Analytical studies \cite{last2001dictionary} can be further divided into three major groups based on the temporal direction in the study. Studies which start with an outcome and look back in time for exposure are known as \textbf{case-control studies}. If the study begins with an exposure and concludes with an outcome,  we refer to them as \textbf{cohort studies} \cite{doll2000smoking,hannaford1998risk,kim2000comparison,huang1999waist}. 
If we only consider a single point in time,  where the outcome and the exposure may both be present at the same time,  we refer to the study as \textbf{cross-sectional} \cite{last2001dictionary}. 

\textbf{Descriptive} study designs, which have no outcome of interest, mostly deal with the frequency and the distribution of risk factors in populations and enable us to assess the extent of a disease of interest. These study designs are usually used to build hypotheses,  thereby building the framework for future clinical research. 

In Figure 1, we categorized studies as analytical or descriptive first and then categorized them further based on temporal direction. Alternatively, and equally correctly, we could have categorized them based on temporal direction first. Studies that start with a cohort selection and follow the cohort forward in time are called \textbf{prospective}; studies that start with the outcome and look backwards in time are \textbf{retrospective}; and studies that take a snapshot of the patient population at a particular point in time are \textbf{cross-sectional}. Within each of these categories, we could further divide them as analytical or descriptive. For example, a case-control study is a retrospective analytical study; retrospective descriptive studies are also possible.

In the following section,  we will discuss the aforementioned study-designs along with clinically relevant examples. 
We will also discuss how certain study designs may introduce biases and confounding factors.

\subsection{Cohort Studies}
Cohort studies are also known as incidence, follow-up, forward-looking, longitudinal, prospective or concurrent studies \cite{lilienfeld1994foundations}. The defining feature of these studies is that we follow a cohort of patients over time. The cohort consists of two kinds of patients: those who are exposed to a particular factor of interest (called the \emph{exposed} group) and those who are not exposed (\emph{unexposed} or \emph{control} group) at a particular point in time (\textit{baseline}). We follow this cohort and compare the incidence of an outcome between exposed group with the unexposed group \cite{beral1999mortality,seman1999lipoprotein,colditz1996risk}.  
Often the exposure of interest is a treatment: the researchers are interested in assessing the effect of a treatment on an outcome.
In this case, the exposed group can also be called the \emph{treatment} group.

As an illustration, let us consider the example of studying the effect of obesity on diabetes. 
First, we decide on a baseline. Baseline can be a particular date (say Jan. 1st, 2005) or an event of interest such as
patients reaching a particular age or developing a particular condition.
At baseline, we take a cross-section of our population.  
Some of our patients already have diabetes, other do not; some of them are obese
some of them are not. Since our interest is the risk of \textit{developing} diabetes, we exclude
all patients who are already diabetic; the remaining patients form our study cohort.  
Our exposure of interest is obesity.
The obese patients form the \textit{exposed} group and the non-obese patients the \textit{unexposed} group. 
We then follow the cohort forward in time and observe how many patients develop diabetes
and how many remain non-diabetic among both the exposed and the unexposed patients. 
We can then compute the odds (or hazard ratio) of developing diabetes among 
those who are obese versus those who are not.

Cohort studies are considered to be the best study designs for ascertaining the incidence, natural progression of a disorder as the temporal relationship between the exposure and the outcome \cite{walline2001designing}. They are also useful in analyzing multiple outcomes that might arise after a single exposure. For example, smoking (i.e. exposure) might lead to multiple outcomes such as stroke, oral cancer and heart disease. 

However, such study designs come with certain caveats. Firstly, selection bias is inherent in such cohort studies \cite{sackett1979bias}. For example, in a cohort study analyzing effects of smoking on T2DM, those who smoke would differ in other important ways (lifestyle) from those who do not smoke. In order to validate the effect of exposure (i.e. smoking), both the cases and controls must be similar in all respects except for the absence/presence of exposure and the outcomes. Secondly, loss of subjects due to censoring might be a possibility, not only when the study is short, but particularly with longitudinal studies that continue for decades. For example, progression from T2DM to associated complications such as Peripheral Vascular Disease (PVD) and Ischemic Heart Disease (IHD) takes around 5 to 10 years, and subjects may drop out over this long period.

\subsection{Case-Control Studies}
Case-Control studies are study designs which look backward: i.e., the study cohort is defined 
at a particular time (e.g. first heart attack) and the study looks backwards in time to analyze the patients' exposure(s). In such studies two groups are compared, one consisting of patients with the outcome in questions (\emph{cases}) and another one consisting of patients without the outcome (\emph{controls}). Case-control study design can be used to identify risk factors that may contribute to an outcome by looking backwards in time and comparing the exposure histories of patients across the case and control groups: exposure that are more prominent in the case group can be risk factors and exposures more prominent in the control group can be protective factors--when the subjects are similar except for the exposure \cite{torrey2000antecedents,xu2011risk,osaki2001factors,avidan2001risk}. 

Let us return to our example (as illustrated in Section 6.1) of estimating the effect of obesity on incident (newly developed) diabetes. We would select a group of patients (cases) newly diagnosed with diabetes and matched them with another group of non-diabetic patients (controls). Matching is performed on important characteristics in an attempt to reduce the differences between the groups to the exposure alone. We would track these patients backwards in time 
for a fixed number of years to determine their exposure history (whether they were obese or not). We can then compute the odds of developing diabetes in obese patients.

Since we start with known outcomes, these study designs are very useful in the investigating (i) slow-onset diseases such as cancer and T2DM where it takes a long time for the disease to develop; and (ii) rare outcomes. In both cases, finding cases--patients who have the outcome (disease) in question--would be difficult prospectively,  either because the study would require a long follow-up time or a very large initial sample size. In case-control studies, cases and 
controls are  identified at the beginning of the study and are followed retrospectively.

Conversely, case-control study designs are inefficient when the exposure rate is low, as researchers would have to analyze the entire data cohort to identify one patient who had the exposure. For example using a case-control study design to investigate the effect of pancreatic cancer (exposure) on T2DM (outcome) would be impractical because the pancreatic cancer is very rare. In such cases, the cohort study design is more effective.

Case-control designs are susceptible to biases. Recall bias can be introduced, where the exposure may occur at a time when the patient is not under study, and thus an exposure may remain undetected.
Specifically, in case-control studies, selection of the control group can also bias the results of the study and therefore researchers should provide clear eligibility criteria for the outcome being studied, such as age, gender, racial makeup or ethnicity.  

In risk prediction, case control study designs are widely used due to their efficiency in detecting the association between risk factors (exposure) and outcome.

\subsection{Cross-Section Studies}
Cross-sectional studies seek a comparison between cases and controls by collecting data at one specific point in time \cite{lee1994odds}. Such study designs differ from case-control and cohort studies in that they aim to make inferences based on data that is collected only once rather than collected at multiple time points \cite{mann2003observational}. 

Returning to our example of measuring the effect of obesity on diabetes, we would take a cross-section
at a well-defined point in time (e.g. Jan 1, 2007). At this time, some patients are diabetic, other are not; and some patients
are obese and other are not. We can compute the odds of having (not developing!) diabetes among 
obese patients versus those who are not obese.

Note that this study design does not require the exposure to precede the outcome, thus cannot
discover causal relationships. For example, some of the obese patients may have already been
diabetic when they became obese.

The key advantage of the cross-sectional study design is that it only requires data at a single
point in time. In risk prediction, cross-sectional study designs are widely used due to their ability to detect the association between risk factors (exposure) and outcome using data at a single point in time.
Strictly speaking, for risk prediction, association is sufficient; the key drawback of this study design
is that it cannot discover causal relationships is neutralized.

\subsection{Descriptive Studies}
Descriptive studies are usually designed to analyze the distribution of variables, without regard to 
an outcome \cite{walline2001designing,hennekens1987epidemiology}. The defining characteristic is that there are no cases or controls\cite{gabert1993recreational,krane1988all,jaremin1996death,marshall1997variation,giordano2002breast,weiss2005epidemiology,anderson1991population}. 

At an individual level, descriptive studies include studies reporting an unusual disease or association and at a population level, they are often used for analyzing the medical state of population for health-care planning \cite{tough2000effects,bider1999incidence,dunn1996recent,steegers1998multiple}. For example, descriptive designs are widely used to investigate tobacco usage within a population, age group, gender or socio-economic class. 

One type of descriptive studies are correlational studies which aim to identify associations between conditions: they typically investigate how predictive
is a condition of another condition. With no information about the temporal ordering of these conditions, we would not consider them as exposure
of outcomes. 
One common use of correlation analysis is to study the relationship between an index disease and its comorbidities; or the relationships among
diseases in multimorbid disease clusters. In this case, the study is referred to as comorbidity analysis.
%
For example, consider our cohort of prediabetic patients, many of whom have different conditions (e.g. high cholesterol, obesity, high blood pressure) at baseline. Using such study designs, we can estimate the prevalence of these comorbid conditions. Further such analysis can lead to future estimation of sequential patterns in which such diseases occur. \\

\section {{Approaches}}\label{sec:Approaches}

Having defined some metrics of interest in Section \ref{sec:Metrics}, and having selected a study design
from Section \ref{sec:StudyDesign} to transform our {EHR} data into an analytics matrix that data mining
methods can operate on, we appear ready to answer the clinical questions we set out to solve.
While the data format may suggest that we can directly apply our existing data mining techniques, 
the data itself creates some challenges, which we described in Section \ref{sec:DataChallenges} in detail. 

In this section, we describe approaches to address the challenges related to censoring, the irregular temporal nature of the data and confounding. 
Approaches are concepts and ideas that provide high-level solutions to these challenges; they are not algorithms per se. 
Within these high-level solutions, concrete data mining techniques can be developed.  
Our focus in this section is to describe these high-level ideas and later, partly in this section as illustrative examples, but mostly in the subsequent sections, we will discuss some concrete analytics techniques that use these approaches.

Several survey articles and books have been published on the approaches we discuss in this section. Chung et al. \cite{chung1991survival} provided a survey of statistical methods, which analyzes the duration of time until an event of interest occurs. Such modes have been widely used to analyze the survival times of various events (e.g. mortality), analyze the time until recidivism and many other applications. In their paper, they summarized the statistical literature on survival analysis. A textbook by Klein and Moeschberger \cite{klein2005survival} provides a comprehensive overview of various techniques used to handle survival and censored data. Dahejia et al. \cite{dehejia2002propensity} discussed causal inference and sample selection bias in non experimental settings. In their paper, they discussed the use of propensity score-matching methods for non-causal studies and discussed several methods by implementing them using data from the National Supported Work experiment.

 \subsection{Handling Censored Data}
 
As discussed in Section 4.1 Censoring occurs when a patient's trajectory is only partially observable. For example, suppose a study is conducted to measure the impact of a diabetes related drug on mortality rate. In such a study, let us assume that the individual withdrew from the study after following the study course for limited duration. In such a scenario, information about patient's vital statistics is only available until the patient was censored. Such data is common in domains such as healthcare and actuarial science. 

 Survival analysis is an area of statistics that deals with censored data. These approaches usually aim to answer questions such as the following: what proportion of the population will survive past a time of observation and what characteristics influence the probability of survival? To answer, the aforementioned clinical questions, techniques are required which can handle censoring, which is frequently present in EHRs. Techniques to handle censored data, can be divided into three major categories: non-parametric, semi-parametric or parametric.  
 
 Non-parametric techniques do not rely on assumptions about the shape or parameters of the distribution of time to event. Examples of such techniques include Kaplan-Meier estimators \cite{kaplan1958nonparametric} and Nelson- Aalen estimators \cite{Cox}. Rihal et al. \cite{rihal2002incidence} used Kaplan-Meir estimators for incidence and prognostic implications of acute renal failure in patients undergoing percutaneous coronary intervention (PCI). Dormandy et al. \cite{dormandy2005secondary} used Kaplan-Meier estimates in their analysis of patients who were diagnosed with T2DM and were at high risk of data and non-fatal myocardial infarction and stroke. Rossing et al. \cite{rossing1996predictors} used Nelson-Aalen estimators for analyzing the predictors of mortality in insulin dependent diabetes. Ekinci et al. \cite{ekinci2011dietary} used such non-parametric techniques for exploring salt intake consumption and mortality in patients with diagnosed with T2DM. 
 
Parametric techniques often rely on theoritical assumptions about the shape or parameters of the distribution of time to event. Examples of such technique are the accelerated failure time models  (AFT models) \cite{keiding1997role} which are an alternative to the widely used proportional hazard models. Babuin et al. \cite{babuin2008elevated} determined whether troponin elevations influence short and long term mortality in medical intensive care unit patients. Wilson et al. used  \cite{wilson2008prediction} used AFT models to predict cardiovascular risk by using predictors such as age, gender,  high-density lipoprotein cholesterol, diabetes mellitus (DM), cholesterol, smoking status, systolic blood pressure and body mass index (BMI). 

Semi-parametric techniques have both parametric and nonparametric components. An example of such a technique is the proportional hazards model. Proportional hazards models relate one or more covariates with the time that passes before some event occurs. Yadav et al. \cite{yadav0231} used the proportional hazards model for risk assessment of comorbid conditions in T2DM. They identified how risks vary across sub-populations for the same outcome. The sub-populations were defined by using diagnosis codes such as hypertension, hyperlipidemia and T2DM with time to death being modeled as the outcome of interest. Martingale residuals were used to compute the risks. Vinzamury and Reddy \cite{ChandanActive} extended proportional hazards regression and proposed sophisticated techniques which can capture grouping and correlation of features effectively. They proposed novel regularization frameworks to handle correlation and sparsity present in EHR data. Further, they demonstrated the applicability of their technique by identifying clinically relevant variables related to heart failure readmission. 
 
 \subsection{{Handling Irregular Time Series Data}} 
 
Data stored in EHRs is usually collected through longitudinal study. In such studies, the subject outcomes, treatments or exposures are collected at multiple follow-up times, usually at irregular intervals. For example, patients diagnosed with T2DM might be followed over time and annual measures such as Hemoglobin A1c and GFR are collected to characterize the disease burden and health status, respectively. As these repeated measures are correlated within the subject, they require sophisticated analysis techniques. In what follows, we describe  techniques that are widely used to handle these repeated measurements. In particular, we cover marginal and conditional models, respectively. These models handle unevenly spaced (irregular) EHRs by assuming a correlation structure among multiple clinical observations of a patient recorded at different time points. 

Marginal models are also known as the population averaged model as they make inferences about population averages. In such models, the target of inference is usually the population and these models are used to describe the effect of covariates on the average response. They are also used to contrast the means in sub-populations that share common covariate values. For example, consider a cohort of pre-diabetic patients with elevated cholesterol levels. In this cohort, if we are interested in estimating the progression of patients to full-blown T2DM, we would probably want to use the population-averaged coefficients. Generalized Estimating equations (GEE's) are mostly used for parameter estimation in marginal models. This approach is computationally straightforward and with care can handle missing data, even when the covariance has been misspecified. Such modeling techniques are widely used in epidemiological studies, particularly in multi-site cohort studies as they incorporate the effect of unmeasured dependence across outcomes.
 
 Conditional models \cite{laird1982random} are also known as the locally averaged models as they usually make inferences about individual subjects. The estimates are based on averaging or smoothing done by the model, but more locally, are based on sources of dependence in estimating model parameters. For example, consider once again our aforementioned cohort of pre-diabetic patients with elevated cholesterol levels. In this cohort, if we are interested in estimating the effect of statin across every individual, we would use conditional models. Yamaoka et al. \cite{yamaoka2005efficacy} used conditional models to evaluate the effect of lifestyle change recommendations for preventing T2DM in individuals at high risk. They observed that lifestyle education intervention reduced glucose levels by 0.84 mmol/l in the case group as compared to the control group.

 \subsection{Handling Confounding }
A confounding variable can be defined as an extraneous variable that correlates with both the dependent and the independent variable. To handle confounding we discuss techniques such as propensity scoring and inverse probability weighing. \\
Statistical matching techniques such as propensity score matching (PSM) \cite{peikes2008propensity} aim to estimate the effect of an intervention by incorporating the effect of covariates that predict receiving the intervention. They aim to reduce the bias caused by confounding variables. PSM creates a group by employing the predicted probability of group membership which is usually obtained from logistic regression. The key advantage of PSM is that by using a linear combination of features, it balances both the intervention and the non-intervention group on a large number of covariates. One disadvantage of PSM is that it only accounts for known covariates i.e. variables which are observed. Another issue is that PSM requires large samples, with substantial similarities in terms of subjects between treatment and control groups. Polkinghorne et al. \cite{polkinghorne2004vascular} used PSM to analyze the inception and intervention rate of native arteriovenous fistula (AVF). \\
 Inverse probability weighting \cite{hogan2004instrumental} is a statistical technique for calculating statistics which aim to standardized to a population as compared to the population from which the data was collected. Instead of adjusting for the propensity score, the subjects are usually weighted. However, factors such as cost, time or ethical concerns might prohibit researchers from directly sampling from the target population. Robinson et al \cite{robinson2011lack} used inverse probability weighting for examining whether lower serum levels of serum 25-hydroxyvitamin are associated with increased risk of developing type 2 diabetes.  \\

\section{{Clinical Data Mining Methodologies }}\label{sec:survMeth}

The discipline of EHR data mining stands at the intersection of epidemiology, biostatistics
and general data mining.  From epidemiology and biostatistics, clinical data mining has borrowed study design,
the methodology that allows us to organize EHR data into a matrix that is amenable 
to the application of data mining algorithms that can correctly answer meaningful clinical questions. 
It has also borrowed basic approaches from biostatistics and epidemiology to address the challenges that EHR data poses, including censoring, analysis of 
irregular time series data and methodologies for causal inference%
\footnote{The roots of causal inference are in computer science,but has been embraced
by epidemiology and biostatistics resulting in the development of the advanced techniques
we described earlier.}. %
In this section, we focus on the contributions of general data mining. 

Traditionally, data mining techniques \cite{tan2006introduction} are broadly categorized as supervised or
unsupervised: supervised methods take an outcome into account, while unsupervised
methods simply learn from the structure of the data. This distinction neatly maps to
study designs: supervised techniques are applicable to analytical studies and
unsupervised techniques to descriptive studies.

The hallmark of EHR data is its temporal nature, suggesting that data mining techniques be further categorized based
on their ability to take time into account. We call a data mining algorithm and its
resulting model \emph{time-aware}, if outcome of interest depends on time; and we call it \emph{time-agnostic}, if it builds a 
model that does not take time into account. 

Although EHR data is inherently temporal, time is not always of relevance.
The clinical question we aim to answer may be \emph{temporal} if time is of relevance
(i.e. time is part of the question) or it may be \emph{atemporal} (not temporal) if
time is not part of the question. Atemporal questions are naturally answered by time-agnostic data mining techniques.
On the other hand, temporal questions can be either answered by time-aware models
or if the question can be transformed into a simpler atemporal question, they can also be solved
using time-agnostic models.  For example, predicting the risk of 30-day mortality after surgery is a temporal
question (time is part of the question) but it can be solved using time-aware models
(e.g. Cox model) or time-agnostic models (e.g. logistic regression). 

\footnotesize
\begin{table}
\begin{tabular}{ p{3cm}|p{5cm}|p{5cm}}
\hline
 & Unsupervised & Supervised \\ \hline 
Time Agnostic & Atemporal Descriptive Studies (8.1.1) &  Cross-Sectional Studies (8.2), Cohort and Case-Control Studies (8.3.1) \\ \hline
Time Aware & Temporal Descriptive Studies (8.1.2) & Cohort and Case-Control Studies (8.3.2) \\ \hline
\end{tabular}
\caption{Structure for Section 8}
\end{table}

\normalsize

The study design dictates whether a question can be temporal or atemporal and it also determines in large part whether any of the challenges posed by {EHR} data can be successfully addressed. For this reason, we describe data mining techniques that are commonly applied in the context of the applicable study designs. In Table 1, we present the structure of the following subsections. 

\subsection{Descriptive Studies}

Descriptive studies represent the broadest variety of inquires we can undertake, ranging from 
simple statistics (prevalence rate, incidence rate) to descriptions of the progression
of a particular diseases via case studies. Such simple applications do not require
data mining, but data mining techniques enable more advanced applications including 
comorbidity analysis and trajectory mining. While descriptive studies cover a wide range of applications, their defining 
characteristic is that no comparison is made between patients (or patient groups)
with and without a particular outcome. Without a particular outcome, we cannot have
outcome labels and hence the problem at hand is \emph{unsupervised}. 

Descriptive studies are commonly used to answer both temporal and atemporal
clinical questions. For example, estimating prevalence rates
at a particular time is an atemporal clinical question, while extracting the 
trajectory of a patient as sequences of diagnosis codes is naturally a temporal clinical
question.  Therefore both time-aware and time-agnostic data mining techniques are
applicable for descriptive studies.

\subsubsection{\textbf{Atemporal Descriptive Studies: }}  Atemporal descriptive studies can arguably be handled using standard textbook methods.
A prototypical application of this nature would be to take a snapshot of the population at a particular time
and cluster the patients based on the conditions they present. One option is
to ``flatten" the temporal dimension of the data through temporal abstraction, e.g. by extracting features and applying non-temporal
unsupervised techniques.

In this section, we discuss how unsupervised techniques have been widely used for \emph{identifying clusters} of patients that have similar characteristics (e.g. demographics, medications, diagnosis codes, laboratory test results) and for finding \emph{associations} between clinical concepts (e.g. medications, diagnosis codes and demographic attributes). Next, we describe how these techniques make use of the approaches we discussed in Section 7. \\ 
\textbf{Clustering:} Gotz et al. \cite{Gotz2011} used clustering techniques for identifying a cohort of patients similar to a patient under observation. They used the cohort as a surrogate for near-term physiological assessment of the target patient. Roque et al. \cite{Roque2011a} stratified patients using hierarchical clustering, where the distance between patient records was computed using the cosine similarity of diagnosis codes. Along similar lines, Doshi et al. \cite{Doshi-Velez2014} investigated the patterns of co-occurring diseases for patients diagnosed with autism spectrum disorders (ASD). They identified multiple ASD related patterns using hierarchical clustering. They further discussed how the aforementioned patterns can be attributed to genetic and environmental factors. Kalankesh et al. \cite{Kalankesh2013} noted that representing the medical state of a patient with diagnosis codes can lead to sparse clusters since EHRs contains a large number of diagnosis codes often running into thousands. To overcome this problem, they used Principal Component Analysis (PCA) \cite{dunteman1989principal} to reduce the dimensionality, thereby making the structure more amenable for visualization and clustering.  Marlin et al. \cite{marlin2012unsupervised} developed a probabilistic clustering method along with temporal abstraction to mitigate the effects of unevenly spaced data, which is inherent in EHRs. \\

\textbf{Association Analysis:} Association rule mining techniques \cite{agrawal1994fast} such as Apriori have also been used on EHR data to identify associations among clinical concepts (medications, laboratory results and problem diagnoses). These techniques have the ability to discover associations and interpretable patterns from EHRs data. However, the performance of such techniques often deteriorates when there are a large number of clinical variables present in EHRs. Wright et al. \cite{Wright2010} used the Apriori framework to detect transitive associations between laboratory test results and diagnosis codes and between laboratory test results and medications. For example, they observed some unexpected associations between hypertension and insulin. They attributed this finding to co-occurring diseases and proposed a novel way to identify such transitive associations. Cao et al. \cite{Cao2005a} used co-occurrence statistics to identify direct and indirect associations among medical concepts. Holmes et al. \cite{Holmes2011a} used statistical approaches to detect associations between rare diseases. They observed that analyzing cohorts comprised of sick patients leads to identification of significant findings. Shin et al. \cite{Shin2010} used association rule mining to identify co-morbidities (e.g. non-insulin dependent diabetes mellitus (NIDDM) and cerebral infarction) which are strongly associated with hypertension. Hanauer et al. \cite{Hanauer2009} used statistical tests to observe common pathways for diseases such as granuloma annulare and osteoarthritis. \\

\textbf{Challenges and Limitations:} In Section 4, we identified a number of challenges, centered around three key challenges: censoring, 
irregular time series, and causation. In this paragraph, we describe the limitations of these methods in relation to these challenges.

While individual techniques all have their own limitations, methods that are applied specifically to atemporal descriptive studies
share some limitations stemming from the nature of the study design.
Atemporal descriptive studies, as all atemporal studies, deal with information collected at one time instant (at \emph{baseline})
or summarized until baseline. With only one time point, this study design completely sidesteps the challenges stemming from irregular time series---naturally at the cost of losing information: a potentially long sequence of information has been aggregated to a single time point.

Atemporal studies (and the data mining methods that are applicable to these studies)
are susceptible to biases stemming from censoring. Right censoring is obviously present as 
we cannot observe any change in the patients' condition after baseline.  
Suppose we have two patients, neither of whom has a condition at baseline,
but one develops it shortly after baseline, while the other develops only decades later (or not at all). These patients are
clearly different for the purpose of the analysis, but the atemporal methodologies consider them identical, which consitutes information loss.
A similar situation arises with left censoring: we may not know how long a patient has had a condition for but if we had
that information, we could obtain more accurate results.

Causation is generally not possible under atemporal descriptive studies. First, descriptive studies do not differentiate between exposures
and outcomes; thus trying to establish causation is meaningless. Second, even if we attempted to establish causation between
two conditions, due to the atemporal nature of the study, we cannot even ascertain that one happens before the other.

While these methods are rather limited, their popularity stems from their simplicity. Most off-the-self data mining algorithms are
directly applicable.

\subsubsection{\textbf{Temporal Descriptive Studies:}} Time plays an important role in the clinical questions. For example, the sequence of events, timing between events, etc. Standard textbook data mining techniques exist to solve such problems (e.g. 
sequence mining, Markov models), but to achieve better results, significant improvements have been proposed. We broadly classify the approaches that can be carried out using such techniques as those which use time-aware techniques, e.g., sequence mining, time-lagged correlations, etc., and those which simplify the problem and apply time-agnosic techniques e.g., temporal-abstraction (summarizing the longitudinal data) and Hidden Markov Models (HMM) trajectory clustering (using HMM to simplify away time so that standard clustering is applicable). \\
\textbf{Temporal Abstraction Framework:} The temporal abstraction framework has been frequently used to prepare patterns from EHR data. Patterns can be abstracted using state representations (e.g. high, medium or low) or trend representations (e.g. increasing, decreasing, constant). Shahar et al.  \cite{Shahar1997} provided a mechanism to abstract patterns from unevenly spaced time-series. Such time-series are common in EHR data elements such as laboratories test results and vitals. They further proposed temporal logic relations to combine patterns generated from univariate time-series. Sacchi et al. \cite{Sacchi2007} extended the temporal abstraction framework to generate temporal association rules (TARs). In TAR\rq s, the antecedent and the consequent both consist of temporal patterns generated using the temporal abstraction framework. Jin et al. \cite{Jin2008} further extended the TAR framework, to generate rules for mining surprising patterns. In such patterns, certain events lead to unexpected outcomes, e.g. taking multiple medicines together sometimes causes an adverse reaction. Batal et al. \cite{batal2009multivariate} used the temporal abstraction framework to propose the Segmented Time Series Feature mining algorithm for identifying the frequent patterns from an unevenly sampled time-series. Such modeling techniques have their own set of challenges. Patterns generated from individual patient time series are susceptible to noise. Further, such patterns can be of uneven temporal duration. \\
\textbf{Trajectory Clustering:} Clustering techniques have also been used to group EHR data. Ghassempour et al. \cite{ghassempour2014clustering} used HMM to cluster patient medical trajectories. In their approach, they used both categorical variables (diagnosis codes) and continuous variables (vitals and laboratories test results) for clustering. They first mapped each medical trajectory to an HMM and then used KL divergence to compute the distance between two HMM's. \\
\textbf{Sequential Rule Mining:} Researchers have explored sequential association rule mining techniques for identifying causal relationships between diagnosis codes. Hanauer and Ramakrishnan \cite{Hanauer} identified  pairs of ICD-9 codes which are highly associated. They observed interesting temporal relationships between hypothyroidism and shingles (herpes reactivation).  Liao and Chen \cite{Liao2013} proposed a sequential pattern mining approach to mine sequences with gap constraints. Such gaps represent the delay between two concepts. Hripsack et al. \cite{Hripcsak2013} measured lagged linear correlation between EHR variables and healthcare process events. In their analysis, they considered five common healthcare process events: inpatient admission, outpatient visit, inpatient discharge, ambulatory surgery and emergency department visit and computed their correlation with several EHR variables such as laboratory values and concepts extracted from clinical notes. \\
\textbf{Visualization:} Research has also been carried out in analytical reasoning facilitated by advanced interactive visual interfaces. Research has been carried out by highlighting the opportunities and associated challenges \cite{caban2015visual}, cohort analysis and exploration \cite{zhang2014iterative,gotz2012icda}, exploring comorbidities \cite{gotz2012multifaceted,sun2010diseaseatlas}, exploring concepts \cite{cao2011solarmap}, clinical decision support \cite{gotz2011visual}, cohort identification \cite{cao2011dicon}, disease network visualization \cite{perer2012matrixflow} and temporal frequent event sequences \cite{perer2014frequence}. 

\textbf{Challenges and Limitations:} 
Method applied in the context of temporal descriptive studies have access to granular temporal information, and thus, in theory,
they can utilize more information than the atemporal descriptive methods from Section 8.1.1.
Even when the methods themselves are unable to utilize the granular temporal information, the study design allows
them to produce results that are qualitatively different. 
Atemporal descriptive studies are limited to using prevalence (proportion of patients in a population 
presenting with a condition), while temporal descriptive studies can use incidence (proportion of patients \textit{developing} 
disease within a relatively short time frame) as well as prevalence.

Naturally, these techniques can overcome the challenges of irregular time series either directly (e.g. time lagged correlation)
or through transforming a temporal problem to an atemporal one by temporal abstraction or by enumerating the sequences of the events.

Another challenge that temporal studies can overcome is handling censored data. The information loss in conjunction with
censoring that we described in the previous section can be reduced. In case of atemporal descriptive studies, we have
absolutely no information regarding when patients develop each condition of interest; the only information we have is
binary indicating whether they have it at baseline. In contrast, in case of temporal descriptive studies, even in the presence of left
censoring we know that a patient has had the condition for at least a certain amount of information and conversely,
in case of right censoring, we know that the patient remains free of that condition for at least a certain amount of time.
While the techniques we described in this section, did not make use of this information, it is possible to develop techniques 
based on survival analysis that can.

Similarly to atemporal descriptive studies, causation remains impractical for temporal descriptive studies.
In case of causation, we are generally interested in the causal effect of an exposure on an outcome.
These methods are still operating under the framework of descriptive studies, thus they do not distinguish between exposure and outcome.

\subsection{Cross-Sectional Design}
Cross-sectional studies are carried out by collecting data at one time point. The aim of such studies is usually to identify and analyze the correlations between risk factors and the outcome of interest. In such studies, data is often collected on individual characteristics, such as exposure to risk factors, demographic attributes and information about the outcome. In what follows, we will describe the techniques often used for such study designs along with examples of research carried out using data mining algorithms.

When a study is designed as cross-sectional, supervised time agnostic data mining techniques are the natural modeling choices. When the study is inherently temporal and  employs a cross-sectional design it can still be solved using supervised time agnostic techniques, but we incur some loss of information due to temporal abstraction. For example, a study investigating the 30-day mortality of patients following an exposure can be modeled using supervised, non-temporal techniques as long as we only consider a binary outcome, namely,  whether the patients survived for 30 days or not. Since interest in a specific outcome is very natural and there is great appeal in simplifying these problems to become solvable through relatively simple supervised non-temporal data mining techniques, such techniques have been applied to a broad spectrum of problems, including risk prediction for hospitalization, re-hospitalization, diagnostic and prognostic reasoning. \\
\textbf{Rule Based Methodologies:} White et al. \cite{White2013} conducted a large scale study for analyzing web search logs for detection of adverse events related to the drug pair, paroxetine and pravastatin. They analyzed whether the drug interaction leads to hyperglycemia. Iyer et al. \cite{Iyer} used NLP techniques for mining clinical notes to identify events related to adverse drug-drug associations. Haerian et al. \cite{Haerian2012} hypothesized that adverse events might be caused by the patient's underlying medical condition. Along similar lines, Vilar et al. \cite{Vilar2012} used disproportionality based techniques to analyze adverse drug events related to pancreatitis, Li et al.\cite{Li} used penalized logistic regression to analyze associations between ADRs and Epstein et al. \cite{Epstein} used NLP techniques to analyze medication and food allergies. Supervised non-temporal methodologies have been frequently used in the form of rule-based techniques for cohort identification. Phenotyping algorithms for diseases such as celiac disease, neuropsychiatric disorders, drug-induced liver injury and T2DM \cite{Pathak2013,Carroll2011,Xu2011} have been widely explored. Supervised pattern mining approaches using the temporal abstraction framework have been used for predicting Heparin Induced Thrombocytopenia (HIT) \cite{batal2009multivariate}. Batal and Hauskrecht \cite{Batal2010d} used such methodologies to generate minimal predictive rules for Heparin Platelet Factor 4 antibody (HPF4) test orders. They further extended their approach by introducing the minimal predictive patterns (MPP) framework wherein they directly mine a set of highly discriminative patterns \cite{Batal2012a}. Those patterns were later used for classification of related tasks. 

\textbf{Bayesian Networks:} Bayesian Networks have also been used to model EHRs for diagnostic reasoning (constructing the medical state of the patient using laboratory test results), prognostic reasoning (prediction about the future) and discovering functional static interactions between the outcome and the predictors \cite{Lucas2004}. Zhao et al. \cite{zhao2011combining} integrated EHR data with knowledge from other sources such as Pubmed to develop a weighted Bayesian network for pancreatic cancer prediction. They also discussed how their approach can be used to detect clinically irrelevant variables for disease prediction. Sverchkov et al. \cite{sverchkov2012multivariate} compared clinical datasets by capturing the clinical relationships between the individual datasets by using the Bayesian networks. The multivariate probability distributions were then used to compare the clinical datasets. 

\textbf{Challenges and Limitations:}  
Methods under cross-sectional design are similar to methods under descriptive atemporal design in that they are all atemporal:
they only consider a cross-section taken at one point in time. Therefore, the limitations associated with atemporal design, namely the lack
of ability to handle censoring and to make use of the granular time information in the EHR apply to cross-sectional designs, as well.

The key difference between  a cross-sectional and descriptive study design lies in the existence of comparison
groups and thus a distinction between exposure and outcome. 
Although this difference allows for qualitatively different results--we can now measure risks--it still is not practical for causal
inference because we cannot establish the temporal relationship between exposure and outcome. To establish causation,
we need to ascertain that the exposure precedes the outcome.
 
\subsection{Cohort and Case-Control Study Design}
Cohort and Case-Control studies compare patient groups with different exposures over time and record their outcomes.  They differ in the direction in which
time is observed: in cohort studies patients are followed from exposure to outcome and in Case-Control studies, patients are followed from outcome to exposures.
While this difference has far-reaching consequences on the required sample sizes, exposure rates and the metrics we can estimate, once the design matrix has been constructed,
the same data mining methods apply to both of these study designs. Hence, we consider these two designs together. \\
What is common across these study designs is that they are best suited  to answer \emph{temporal} questions. As it is typical with temporal questions, we can use either time-aware models or we can simplify the question such that it can be answered using time-agnostic models. In the following paragraphs, we provide examples of both.
\subsubsection{\textbf{Time-Agnostic Models for Cohort and Case-Control Studies:}}\textbf{Time Agnostic Regression:} 
 Supervised time-agnostic models are commonly employed when time-to-event can be removed
from the clinical question.  For example, time-to-rehospitalization can be simplified to the binary outcome of 30-day rehospitlaization (yes/no)
 of 30-day-rehospitalization (yes/no) which does not include time. 
Applications of supervised time-agnostic modeling include supervised feature creation \cite{aliferis2010local}, predicting the onset of neonatal sepsis \cite{Mani}, potentially preventable events \cite{sarkar2013impact}, 30 day hospital readmissions \cite{Cholleti2012,park2014hierarchical}, post-hospitalization VTE risk \cite{Kawaler2012}, T2DM risk forecasting \cite{Mani2012}, atrial fibrillation \cite{Karnik2012}, 5 year long life expectancy risk calculation \cite{Mathias2013}, risk of depression using diagnosis codes \cite{Huang2014}, survival of heart-lung transplant patients \cite{Oztekin2009}, breast cancer survivability \cite{Sarvestani2010}, 30 day mortality in patients suffering with cardio-vascular diseases, risk of retinopathy in patients suffering from type 1 diabetes mellitus (T1DM) \cite{Skevofilakas2010}, mortality in patients suffering from acute kidney injury \cite{Matheny}, mortality prediction in ICU \cite{Herasevich2013} and risk of dementia \cite{Maroco2011}.  For these analyses, almost all flavors of common predictive modeling techniques ( decision trees \cite{Mani2012,Sarvestani2010,Austin2012a}, ensemble techniques (e.g. bagging,boosting,random forests) \cite{Cholleti2012,Kawaler2012,Mani,Karnik2012},naive Bayes \cite{Kawaler2012,Karnik2012,Sarvestani2010}, linear regression, support vector machines \cite{Mani2012} and logistic regression \cite{Zhai2014,Mani2012,Cholleti2012,Huang2014,Chang2011}  have been used. These techniques have also been used for identification of regional differences in breast cancer survival rates despite guidelines \cite{Ito2009}, comparison of cancer survival rates across continents \cite{Coleman2008}, comparison of cancer and survival patients over time, exploring relationships between hospital surgical volumes and 5 year relationship of stomach cancers \cite{Nomura2003}, comparing dosage volumes of warfarin in European-American and African-American \cite{Ramirez2012}, postpartum depression rates in Asian-American subgroups (Indian, Vietnamese, Chinese, Korean, Filipino, Japanese,) \cite{Goyal}, analyzing the effect of different ethnicities on different levels of susceptibility to diabetes related complications and studying the detrimental effect of fibrates on women as compared to men in a population presenting with high cholesterol levels. 

Ghalwash et al. \cite{ghalwash2014data} proposed predictive modeling technique to find a suitable duration of the hemoadsorption (HA) therapy control and observed that their method led to substantial monetary savings. Sun et al. \cite{sun2014predicting} worked on predicting the risk and timing of deterioration in hypertension control by analyzing critical points in time, at which hypertension status is at borderline (clinical limit separating control vs out-of-control)l. Wang et al. \cite{wang2015dynamic} developed a dynamic Poisson autoregressive model for flu forecasting where in they allowed the autoregressive model to change over time. Panahiazar et al. \cite{panahiazar2015using} built a heart failure risk prediction model using several machine learning techniques where in they included multiple comorbidities which lead to improvement in prognostic predictive accuracy. Wang et al. \cite{wang2014clinical} proposed Multilinear sparse logistic regression to handle data in the form of multi-dimensional arrays. They used their methods to predict the onset risk of patients with Alzheimer's risk and heart failure. \\
\textbf{Overcoming Challenges:} Such techniques also have their own share of caveats. Causal analysis is not possible as time-to-event data is often ruled out and there is no way to ascertain the relationship between diseases and the outcome of interest. The inherent design of such techniques rules out longitudinal analysis. Temporal abstraction is also employed to summarize time. As comparison groups are available in such study designs they are well-suited for applications such as risk prediction. Further, handling right censored data is not possible but handling left censored data and interval censored data is plausible.

\textbf{Challenges and Limitations:} Case-control and cohort study designs are the most flexible and informative designs among the study designs we consider. These designs have dedicated exposures and outcomes, allowing for measuring risks; and they have detailed temporal information
allowing to measure time to event and establish the precedence (temporal ordering) of events.
Unfortunately the time-agnostic methods handicap the study design. Simplifying this study design to become atemporal, effectively 
renders the study design into a cross-sectional design with some amount of extra information about outcome (did it happen within 
a certain time-frame). Since the design has been effectively reduced to cross-sectional, the limitations of cross-section designs stand, but
to a lesser degree.

In case of a cross-sectional design, we have no information about the timing of the outcome. In contrast, in case of case-control studies,
even when we use time-agnostic methods, we usually have some information about the time of the outcome: e.g. the patient did not
develop the disease in a certain time frame (say 5 years). The information these methods discard concerns patient who did not develop the disease: patients who got censored (are lost to follow-up within 5 years without developing the disease) are ignored, even though
these patients carry partial information to the effect that they did not develop the disease for some amount of time (which is less than 5 years).

The key advantage of these methods over cross-sectional design lies in causation.
Although we do not have detailed time-to-event data, we know that the exposure precedes the outcome, and we also have
a distinction between exposure and outcome, thus causal inference is possible.

\subsubsection{\textbf{Time-Aware Models for Cohort and Case/Control Studies:}}
Supervised time-aware models are utilized when the clinical question cannot be simplified or if the simplification to time-agnostic modeling comes at a significant loss of information. 
Such question focuses on the time-to-event itself (clearly cannot be simplified), sequences of events or when time-to-event carries additional information about the outcome.  Continuing with the example of 30-day rehospitalization, by simplifying the outcome to binary yes/no, we lose information since we ignore whether the patient was re-hospitalized in (say) 7 days vs 20 days.  The former case is clearly more severe. 

Many of the temporal clinical questions are related to right censoring. Survival modeling, which was specifically developed for this purpose, is the quintessential technique for this study design.
Survival modeling is a suite of techniques with various specializations that share a common  characteristic of being able to handle time and censoring. Other techniques which incorporate the effect of time include dynamic Bayesian networks, sequential pattern mining, etc. \\
 \textbf{Survival Modeling:} Wells et al. \cite{wells2008predicting} hypothesized that patients diagnosed with T2DM have an increased risk of mortality. They used Cox proportional hazards regression with time to death as the outcome. They also observed that certain interaction terms involving medications and age were significant indicators. Vinzamury and Reddy \cite{ChandanCox} extended Cox proportional hazards regression which aims to capture the grouping and correlation of features effectively. They proposed novel regularization frameworks to handle the correlation and sparsity present in EHR data. They demonstrated the applicability of their technique by identifying clinically relevant variables related to heart failure readmission. Vinzamury et al. \cite{ChandanActive} proposed a novel active learning based survival model wherein continuous feedback from a domain expert can be utilized to refine the model. Survival modeling techniques on time-to-event data have been explored widely in the past. Cox regression \cite{Cox,ChandanCox} is one of the most commonly used survival regression models. Its formulation, namely its semi-parametric nature,with the mild assumption of the proportionality of hazards, makes it ideal for many practical applications in fields such as economics \cite{Coxapp5}, healthcare \cite{Coxapp2,Coxapp3,Coxapp4} and recommendation systems \cite{Coxapp6}. 

Cox models, as most other regression techniques, are susceptible to overfitting. Standard regularization techniques, developed for other regression methods, have
been applied to Cox models, as well. Lasso \cite{lasso} and  elastic-net regularized Cox models \cite{Simon} have been developed, and have been further
extended by regularizing them with convex combinations of L1 and L2 penalties \cite{adapCox}.  We are not aware of regularization for time-dependent covariate 
Cox models \cite{TimeVariantCox}, which would be a straightforward extension. 

Reddy et al. \cite{ChandanActive} proposed a new survival modeling algorithm which uses a sophisticated discriminative gradient based sampling scheme and observed better sampling rates as compared to other sampling strategies. To handle correlated and grouped features, they proposed correlation based regularizers with Cox regression which are commonly seen in many practical problems \cite{ChandanCox}. Kuang et al. \cite{kuangCox} proposed Net-Cox, a network based Cox regression model to handle the high-dimensionality of high-throughput genomic data. They further applied their model to a large-scale survival analysis across multiple ovarian cancer datasets. 

Support vector machines \cite{svm} models have also been extended to handle censored data \cite{svmCox,svmCox1,svmCox2,svmCox3,shiao2014learning}. In such techniques, often the task is converted into a ranking problem via the concordance index. This in turn is efficiently solved using convex optimization techniques. Along similar lines, Khosla et al. \cite{Coxapp1} proposed algorithms which combine margin-based classifiers along with censored regression algorithms to achieve higher accuracies (concordance in this case). They used their technique to identify potential novel risk markers for cardiac problems. 

Research has also been carried out on extending decision trees to handle censored data \cite{treeCox}.  Ishwaran et al. \cite{rsfCox} proposed Random Survival Forests for analyzing right censored survival data. They analyzed splitting rules for growing survival trees, introduced a new measure of mortality and applied it for patients diagnosed with coronary artery disease. Neural nets have also been adapted to handle censored data with varying results \cite{neuCox,neuCox1}. Techniques such as reverse survival \cite{yadav2015forensic} have also been explored in the past wherein they go further back in time. \\
\textbf{Dynamic Bayes Networks:}  While survival models are by far the predominant type of models, other methods that can incorporate temporal information also exist. Dynamic Bayesian networks (DBN) \cite{melnyk2013detection} have been used to model temporal relationships among EHR variables \cite{rana2015predictive}. Nachimuthu et al. \cite{Nachimuthu2010} used DBN's to model temporal relationships between insulin and glucose homeostasis. The modeling was further used to predict the future glucose levels of a patient admitted in an ICU. They also discussed the reasons for using first-order Markov models to model the temporal relationships. Sandri et al. \cite{Sandri2014} used DBNs with multiple order dependencies to impose restrictions on the causal structure, while modeling organ failure in patients admitted to an ICU. In their model, each time-stamp represented a day. They further imposed several constraints such as that no patient discharges were recorded on the second day and that all patients were either deceased or considered discharged on their seventh day. Such constraints were imposed to reduce the complexity of the model. Along similar lines, Rose et al.  \cite{Rose2005a} used DBN\rq s to assist physicians in monitoring the weight of patients suffering from chronic renal failure, Gatti et al. \cite{Gatti2011} used it to model heart failure and Peelen et al. \cite{Peelen2010} used hierarchical DBN\rq s for modeling organ failure. Expectation-Maximization was used to learn conditional probabilities in these DBN\rq s.  \\
\textbf{Sequential Pattern Mining:} In the realm of supervised temporal pattern mining, research has extended the temporal abstraction framework by mining recent temporal patterns for monitoring and event detection problems in patients suffering from T2DM \cite{Batal2012a}. Sengupta et al. \cite{Sengupta2013} used similar techniques for detecting sequential rules associated with the early identification of brain tumors. Simon et. al. \cite{Simon2013} proposed survival association rule mining (SARM) techniques which uses survival modeling techniques to incorporate the effects of dosage and other confounders such as age and gender. \\ 
\textbf{Deep Learning:} Deep Learning has been widely used in conjunction with EHRs for identification of novel phenotypes and robust clinical decision support systems. Lasko et al. \cite{lasko2013computational} used deep learning for phenotype discovery in clinical data. In their analysis, they used a deep learning architecture in conjunction with Guassian process regression to generate phenotypic features that identified multiple population subtypes thereby distinguising the uric-acid signatures of acute leukemia vs gout. They further observed that the phenotypic features were as accurate as gold standard features created by domain experts. Liang et al. \cite{liang2014deep} hypothesized that creating effcient feature representations requires massive manual efforts. To overcome this, they used deep learning based architectures which can express different complex concept levels with multiple layer networks. They used deep belief networks for unsupervised feature extraction. Extracted features were then used to perform supervised learning via SVMs. They observed that their results deliver the promise associated with deep learning. Miotto et al.  \cite{miotto2016deep} proposed a novel unsupervised feature learning algorithm using three layer stack of autoencoders that faciliate predictive modeling for various diseases such as T2DM, cancer and schizophrenia. They observed increased predictive performance when compared with raw EHR data and traditional feature engineering techniques. Choi et al. \cite{choi2015doctor} used Recurrent Neural Networks to predict the medication and diagnosis classes for the next visit using longitudinal data consisting of 260K patients. They further validated their study by validating their results using another cohort and observed how deep learning based architectures can be used to achieve better accuracy for noisy and missing clinical data.\\

\textbf{Challenges and Limitaitons:} These techniques are by far the most successful in terms of overcoming EHRs related challenges. Right, left and interval based censoring can be easily handled by employing techniques such as Cox proportional hazards regression and accelerated failure models. 

The biggest claim of such techniques is their ability to handle causation. As these techniques have comparison groups (i.e. case and control) and can handle time-to-event data, causal analysis can be performed with ease. Further, causation by adjusting for measured confounders can also be analyzed by using marginal structural models and structured nested models. However the literature of such techniques in computer science is very sparse. One area, where more work should be done is to handle unmeasured confounders for the disease of interest. Similarly more research needs to be focused in areas where the effects of confounders need to be adjusted for time-to-event data. \\ 

\section{{Discussion}}\label{sec:Discussion}

\begin{table}[t]
\scriptsize

\begin{longtable}{ |p{4.3cm}|p{9cm}|}

      \hline
\multicolumn{2}{|c|} {  \textbf{Understanding the Natural History of Disease} } \\   \hline
 Descriptive Atemporal & \cite{dunteman1989principal,zhang2002association,Cao2005a,Holmes2011a,zhang2014towards,Hanauer2009,hanauer2013describing,gotz2016adaptive,gotz2014decisionflow,west2015innovative,monsen2010discovering,monsen2015factors} \\  \hline
 Descriptive Temporal & \cite{Hanauer,Liao2013,Hripcsak2013,Sacchi2007,Jin2008,batal2009multivariate,patnaik2011experiences,munson2014data} \\  \hline
 Cross-Sectional &  \cite{dasgupta2014disease,albers2010statistical,albers2012using} \\  \hline
 Time Agnostic Case-Control or Cohort & {}\\  \hline
  Time Aware Case-Control or Cohort &  \cite{Nachimuthu2010,Verduijn2007,Batal2012a,choi2016multi,perotte2013temporal} \\  \hline

  \multicolumn{2}{|c|} { \textbf{Cohort Identification} }\\ \hline 
  Descriptive Atemporal &\cite{Roque2011a,Bauer-Mehren2013}  \\ \hline 
  Descriptive Temporal &\cite{gotz2014methodology,schulam2015clustering,lasko2013computational,wang2015rubik}  \\ \hline 
  Cross-Sectional &\cite{wang2013exploring,peissig2012importance,pathak2012applying,nadkarni2014development,chen2015building,chen2016patient,hripcsak1997automated,hripcsak2013next}  \\ \hline 
  Time Agnostic Case-Control or Cohort & \cite{Carroll2011,Xu2011,Sarvestani2010}  \\ \hline 
  Time Aware Case-Control or Cohort &  \cite{albers2014dynamical,liu2015temporal,che2015deep,albers2014dynamical}  \\ \hline 
  
\multicolumn{2}{|c|} { \textbf{ Risk Prediction/Biomarker Discovery} }\\ \hline  
Descriptive Atemporal & \cite{Gotz2011,vellanki2014nonparametric,miotto2016deep} \\ \hline
 Descriptive Temporal & {}\\ \hline
 Cross-Sectional & \cite{wilcox2003role,sarkar2012improved,letham2013interpretable,ebadollahi2010predicting,feldman2014admission,stiglic2015comprehensible,ngufor2015heterogeneous,byrd2014automatic,liang2014deep}  \\ \hline
 Time Agnostic Case-Control or Cohort &\cite{lin2008exploiting,yadav2016interrogation,westra2011interpretable,Skevofilakas2010,Oztekin2009,ghalwash2014data,somanchi2015early,li2016distributed,li2016regularized1,li2016regularized2,chengrisk,Collins2011,Simon2011,Mani,Matheny,Pakhomov2011,fung2008privacy,wells2008predicting,ChandanActive,li2015constrained,VanderHeijden2014,jackson2011data,Paxton2013,zhao2011combining,Maroco2011,Breault,wang2015dynamic,sun2014predicting} \\ \hline
Time Aware Case-Control or Cohort &   \cite{wang2014unsupervised,ghassempour2014clustering,yadav0231,Gatti2011,Sandri2014,ChandanCox,Peelen2010,choidoctor,wang2013framework,luo2016tensor,hripcsak2015parameterizing,hagar2014survival}  \\ \hline

  \multicolumn{2}{|c|} { \textbf{ Quantifying the effect of Intervention }  }\\ \hline 
 Descriptive Atemporal &  {}\\ \hline 
 Descriptive Temporal &  {}\\ \hline 
 Cross-Sectional &   {}\\ \hline 
 Time Agnostic Case-Control or Cohort & \cite{Schrom2013a,westra2017secondary,yadav2016causal}\\ \hline 
 Time Aware Case-Control or Cohort & {}  \\ \hline 
    
 \multicolumn{2}{|c|} { \textbf{Constructing evidence based guidelines } }\\ \hline 
 Descriptive Atemporal &  {}\\ \hline 
 Descriptive Temporal & \cite{pivovarov2014temporal} \\ \hline 
 Cross-Sectional & {}\\ \hline 
 Time Agnostic Case-Control or Cohort & {mani2007learning,pruinelli2016data}\\ \hline 
 Time Aware Case-Control or Cohort & {}\\ \hline 
 
  \multicolumn{2}{|c|} { \textbf{Adverse Event Detection} }\\ \hline 
    Descriptive Atemporal & {} \\ \hline 
    Descriptive Temporal &   {} \\ \hline 
    Cross-Sectional &  \cite{sathyanarayana2014clinical,White2013,Iyer,Haerian2012,Pathak2013,Carroll2011,Xu2011,bobo2014electronic,zhang2015label,melton2005automated} \\ \hline 
    Time Agnostic Case-Control or Cohort & {} \\ \hline 
    Time Aware Case-Control or Cohort &  {hauskrecht2013outlie}  \\ \hline 
\caption{Data Mining Research in EHRs}
\end{longtable}
\normalsize
\end{table}

Despite its infancy, the health care data mining literature is very rich.
Table 2,  provides a succinct representation of the major research carried out using data mining techniques in conjunction with EHRs. 
For every major application area, presented in section 2, it lists publications related to the various methodologies from 
Section \ref{sec:survMeth}. In the rest of this section, we will explore, discuss and present novel insights about how data mining techniques have been utilized for EHRs. The first three sub-sections correspond to three viewpoints that Table 2 can be viewed from: we can view from
the perspective of the application, the study design and the data mining methodology.
In the fourth and last sub-section, we will discuss what we believe is the most important barrier to the wide-spread use of
data mining in clinical practice.

\subsection{Applications}
The most popular application for healthcare data mining is risk prediction, followed by research aimed at understanding the  natural history of disease.

The goal of risk prediction is to compute the probability of a patient's progression to an outcome (e.g. T2DM) of interest. The reason for its popularity is that it is simply the most natural and immediately impactful application. With numerous data mining and statistical tools and techniques  readily available, such analyses can be performed with ease. The literature using such analyses is rich, providing researchers with opportunities to compare their findings.  

Another popular application area is understanding the natural history of the disease. 
Again, understanding the prevalence, incidence and coincidence of diseases is 
the foundation on which policy decisions can be made and thus disciplines like epidemiology
have spent considerable effort on this application. Also, off-the-shelf statistical or data mining
software, make these studies accessible to a wide range of researchers.

Conversely, certain application areas are virtually unexplored. We found three major reasons for this:
1) these applications may be technically difficult, requiring knowledge of concepts that are not commonly known in data mining; 
2) they require extensive collaborations; or 3) the applications in question are simply not practical.

Quantifying the effect of interventions embodies the first reason: it is technically difficult.
It requires two key technical elements that are not usually part of the
standard data mining toolbox: study design and causal inference.
On the surface, quantifying the effect of interventions, say the effect of a drug on an outcome, 
is a simple causal inference problem that we described earlier in the survey and
some of the solutions, such as propensity score matching, appear directly applicable.
However, a number of epidemiological considerations are necessary to arrive at a valid conclusion.
First, we need to consider the comparison groups. In a population, there are two groups of patients:
those who are subject to the intervention and those who are not. Clearly, patients who are subject to the intervention
form our \textit{exposed} group, but who are the controls (the \textit{unexposed} group)?
Patients who are not subject to the intervention are of many kinds: patients with no intervention at
all; patients who require a weaker intervention; patients who are subject to an alternative, equally effective
intervention; and patients who have already progressed and require a stronger intervention. 
These groups could all serve as controls, but depending on which group we choose, we answer a different clinical question.
For example, if we are interested in the reduction of mortality as a result of starting patients on a first-line drug \textit{earlier},
then our control group should consist of patients who are not taking any drugs for the same disease; if we are interested in the
reduction of side effect by using a particular drug over the standard treatment, then the control group should consist of
patients who take an equally effective alternative drug.
Propensity score matching alone does \textit{not} prevent us from selecting patients from the wrong control 
group, because comparable patients may exist in all of the above groups.

Second, how much exposure is sufficient? For many interventions, relatively long exposure periods are
needed to achieve the desired effect. Requiring a certain exposure period, e.g. we only consider patients
with 180 days of exposure, introduces bias (known as immortality bias), because patients who did not tolerate the
drug or died within 180 days are not included, potentially leading to an overly optimistic estimate for the
effect of the intervention.

The second reason for the unpopularity of certain applications is their need for extensive collaborations.
For example, constructing (or even just evaluating) evidence based guidelines
requires more than just a study design and causal inference; it is decidedly interdisciplinary often critically depending on
interpretation and feedback from researchers with a broad spectrum of clinical expertise.
To give a concrete example, consider the National Comprehensive Cancer Network guidelines for metastatic colon cancer:
"Evidence increasingly suggests that [...] mutation makes response to [...], as single agents or in combination with cytotoxic chemotherapy, highly unlikely." This guideline neither requires nor prohibits the use of certain agents when a specific mutation is present; it merely draws
attention to the possibility that the agent may not work. Such guidelines leave a lot of room for interpretation: do we
administer the referenced drug or select a different one?
Another example is the Surviving Sepsis Campaign 3-hr bundle. This bundle gives a list of actions, such as "administer
broad spectrum antibiotics", that needs to be carried out within 3 hours after the suspicion of sepsis. This guideline
is prescriptive (we know exactly what needs to be done and when), but even this guideline leaves room for interpretation
through the "suspicion of sepsis" phrase: we need to estimate when a well-trained clinical would "suspect" sepsis.

The third reason for applications being unpopular is that they are unnatural. Examples include the usage of descriptive data mining techniques for predicting complications, quantifying the effect of interventions and adverse event detection. Patients are already grouped into cases and controls and therefore case-control, retrospective or cross-sectional studies design would be a much more natural
choice.


\subsection{Study Designs}
Substantially more work has been carried out in retrospective or case-control settings as compared to the descriptive setting. This stems from the nature of research in the medical domain, as research in medical sciences has hitherto been driven by pre-defined outcomes. The defining difference between case-control and descriptive designs is the existence of an outcome. 

The outcome provides a focus to the case/control studies, which does not exist in descriptive
studies. The lack of an outcome combined with the high dimensionality and the associated heterogeneity of EHR data often lead to increased complexity, which translates to exponentially
large number of patterns. If the researchers tune their algorithm to find a small number
of patterns, they tend to be trivial patterns; if they extract a large number of patterns, those patterns
are often redundant and difficult to interpret. Developing algorithms to directly discover
novel patterns is an avenue for method development.

A well-chosen study design with appropriate comparison groups can help address this problem. 
Consider, for example, the difference between the retrospective case-control and cohort study.
In case of the case-control design, we select cases (patients with the outcome in question) and controls (patients without 
the outcome in question) at the end of the study and follow them backwards in time examining their exposures.
If the outcome is rare, this approach helps focus on a relatively small population. 
Exposures associated with the outcome are likely to be present in this smaller population in sufficient amounts,
while "random" (unassociated) outcomes are reduced.  With reduced "random" exposures, we can substantially
reduce the number of uninteresting patterns.
Conversely, if the exposure is rare (relative to the outcome), we would select a cohort study, where
patients are selected based on their exposures at baseline and followed forward in time to see whether they
develop the outcome in question. The rareness of the exposure will control the number of spurious patterns.

\subsection{Methodologies}
Little research has been done to utilize the temporality associated with EHR data. For example, descriptive atemporal studies are more frequently conducted as compared to descriptive temporal techniques. The case is similar with retrospective or case-control studies. We identify a couple of reasons for this phenomenon. First, the duration of EHR data available with healthcare providers rarely exceeds couple of years. Diseases such as T2DM take around 5-10 years for patients to progress from one state to a state of advanced complication. With such a limited duration of data available, this progression cannot be studied effectively. Secondly, censoring and irregular EHR data limits the application of techniques to EHR data often requiring sophisticated techniques (such as those from Section 7.2) and rigorous study designs. 

The prevalent method of representing irregular time series data is to add multiple observations per patient to the observation matrix.
Doing so violates the i.i.d. (independent and identically distributed) assumption of observations, making them correlated.
Section 7.2. describes a number of techniques for addressing this correlation, but a careful look at these techniques reveals that
they are all regression models.  The most popular data mining techniques (SVM, random forests, neural networks) 
are not capable of handling this situation satisfactorily without modification.  
Simply ignoring the correlation could result in reasonable estimates if the number
of observations are similar across the population. In health care, however, this is rarely the case: sick patients contribute much
more data (more frequent visits, more tests per visit) than relatively healthy patients. Ignoring correlation among observations here will
lead to biased estimates (a form of sampling bias).

To make matters worse, many of these techniques use validation (e.g. leave-out validation) to 
avoid or reduce overfitting: performance on a leave-out validation set often serves as the stopping criterion for tree induction. 
The sampling unit for these techniques is typically each row of the observation matrix.  When the rows are observations of a number of patients, the sampling unit has to be the patient (not the observation). Therefore, all observations of the
same patient must either be in the training set or the validation set; we cannot have some observations in the training set and others in the
validation set. Failing to follow this rule can lead to overfitting.

\subsection{Barriers to Clinical Data Mining}
This survey is a testament to the effort that the data mining community has expended on mining EHR data, yet the translation of these results 
into clinical practice is lagging.  This lag is a direct consequence of what the authors of this survey view as the key difference between 
clinical data mining and data mining in general: clinical data mining models have to be validated to a standard that is much higher 
than in many other fields.
Many fields, recommendation systems for example, "only" require that we demonstrate a reproducible performance advantage of the proposed
model over the models in use. This standard of validation can be satisfied relatively easily through textbook validation techniques such as
cross-validation, leave-out validation, etc. Validity in health care often presumes that the model captures some knowledge of human 
physiology and pathophysiology.  Without using predictors rooted in (patho-)physiology for clinical prediction, we run the risk of
capturing a factor that may change at any time, invalidating a model that appears to produce computationally reproducible results.  
A prototypical example of this phenomenon is the Google Flu Trend that used queries related to influenza originating from a
particular geographic region to assess the level of exposure to influenza in that region. As a result of changes in user query 
behavior, the model became invalid and this service is no longer offered.

With recent advances in machine learning (e.g. deep learning), understanding a model's decision process is becoming increasingly difficult.  
As the recent Workshop on Human Interpretability of Machine Learning (ICML 2016) shows, health care is not the only field that 
desires an understanding of the decision process of the machine learned models. A paper by Lipton \cite{lipton2016mythos} draws a parallel between model interpretability and trusthworthiness and describes several criteria for a trustworthy model. Constructing trustworthy models is one direction that can help remove barriers
from the implementation of machine learned models in clinical practice.

\section{Concluding Remarks}

The current landscape of health care and the drivers that shape it virtually ensure that mining EHR data will play an increasingly important role in the future. Major examples of these drivers 
include the transition from the current reimbursement-based health care model to the Accountable Care
Organization model; personalizing care to make it safer, more efficient and to reduce waste; 
and shared decision making, patients' desire to become more involved in their own care.
All of these drivers require strong analytics based on large populations to be successful. 

 The cornerstone of modern medicine is clinical evidence
and generating hypotheses for clinical evidence is one role that data mining is likely to play.
This role is not a replacement of clinical trials but a synergistic role, where data mining
can create high-quality hypotheses that can be validated through clinical trials.
Clinical trials are expensive, thus the number of patients participating in a study is
kept to the minimum required to validate the hypothesis.
Secondary use of such small data sets for hypothesis generation is impractical, creating an opportunity for mining EHR data. 

Even before the broad availability of EHR-based clinical data,
large-scale observational studies over entire populations have been carried out. However, these studies have typically utilized claims data, which exhibit many 
of the key characteristics of EHR data, including the  intermittent generation of longitudinal data, censoring, confounding and the need for a 
robust study design.  Methods utilized for these studies came from biostatistics and epidemiology. The integration of biostatistics techniques into clinical data mining has already started and novel extensions to biostatistics techniques using data mining are being proposed. Similar, interchange of ideas with epidemiology needs to happen but is in an early stage. 

EHR data mining to reach its full potential needs to develop them further in many directions. One of these directions is data representation. EHR data is rich, it consists of highly heterogeneous data, collected from a wide range of sources
(structured and unstructured clinical data, images, omics data, wearable sensors, mobile health) potentially longitudinally
at varying frequencies and resolution. The emergence of such data type heterogeneity necessitates that
we revisit even fundamental questions like how data is best represented for modeling purposes. Other directions include analyzing temporal and sequence data,
handling missing data, and causal inference. Both data mining and biostatistics/epidemiology have
methods to address these issues but they need to be further developed to suit EHR data better. 

The characteristics of EHR data that drive the development of new data mining techniques are not 
unique to EHR data. Heterogeneity is present and poses challenges in many areas, including
earth science and climate; time-to-event data finds its origins in failure analysis; and censoring along with
the intermittent interactions with customers also happens in customer relationship management 
and recommendation systems. Mirroring how techniques for censored data and causal inference are being adopted from
biostatistics and epidemiology into clinical data mining, new developments in EHR mining will likely 
find applications in many other data mining and analytics domains.\\

\bibliographystyle{ACM-Reference-Format-Journals}
{\bibliography{acmsmall-sample-bibfile}

\end{document}


\markboth{Yadav et al.}{ Supplement To Mining Electronic Health Records (EHR): A Survey }

\title{Supplement To Mining Electronic Health Records (EHR): A Survey}
\author{Pranjul Yadav
\affil{ University of Minnesota - Twin Cities}
Michael Steinbach
\affil{ University of Minnesota - Twin Cities}
Vipin Kumar
\affil{University of Minnesota - Twin Cities}
Gyorgy Simon
\affil{University of Minnesota - Twin Cities}
}

\begin{bottomstuff}

\end{bottomstuff}

\maketitle

This online supplement (addendum to the main survey) consists of the research related with various clinical data mining applications as introduced in main survey (Section 2). \\

\section{{Clinical Data Mining Applications}}\label{sec:survApp}

  \subsection{Understanding the Natural History of Disease}
 \cite{jensen2014temporal,teno2001dying,murtagh2008illness,wang2014unsupervised,Roque2011a,Doshi-Velez2014,Wright2010a,Cao2005a,Holmes2011a,Shin2010,dasgupta2014disease}

  \subsection{{Cohort Identification}}
 \cite{friedlin2010comparing,liao2010electronic,liao2010electronic,halpern2014using,meystre2009clinical,kandula2011bootstrapping,rasmussen2014design,castelli1977hdl,newton2013validation,overby2013collaborative,pathak2013electronic,schram2014maastricht,boland2013defining,gotz2014methodology,wang2013exploring,peissig2012importance,pathak2012applying,ho2014limestone,ho2014marble,ho2014extracting,schulam2015clustering,hu2012healthcare}

 \subsection{Risk Prediction/Biomarker Discovery}
 \cite{greenland2004coronary,yadav2016interrogation,knaus1991apache,yadav2015forensic,sarkar2012improved,letham2013interpretable,ebadollahi2010predicting,feldman2014admission,ngufor2015heterogeneous,byrd2014automatic,kamkar2015stable,tran2014framework,lakshmanan2013investigating,vijayakrishnan2014prevalence,Schrom2013a,Simon2013,Harpaz2013,vellanki2014nonparametric,yadav0231,zhao2011combining,algar2003predicting,detsky1986predicting,finks2011predicting,mccrory1993predicting,hay2004predicting,mccarthy2008predicting,epstein1993predicting,ferguson2003comparison,ferguson2002preoperative,ferguson1988diffusing,soohoo2006factors,benz2001predicting,ozkalkanli2009comparison,propst2000complications,klastersky2006outpatient,liu2014clinical,liu2013clinical,liu2013modeling,liu2013sparse,panahiazar2014using} 

\subsection{Quantifying the effect of Intervention}
 \cite{Schrom2013a,campbell1996quantifying,westra2017secondary,prochaska2008methods,ronsmans2011quantifying,law2003quantifying,pruinelli2016data,yadav2016causal} 

\subsection{Constructing Evidence Based Guidelines }
\cite{Liu2012,Eibling2014,agrawal2009clinical,zeng2014systematic,Heselmans2013,Bidra2014,Kinnunen-Amoroso2013,Goh2014,pivovarov2014temporal} 

 \subsection{Adverse Event Detection}
 \cite{Vilar2012,Harpaz2013,Epstein,Iyer,Haerian2012,White2013,bobo2014electronic,pathak2013mining,shen2014using,Nachimuthu2010,Rose2005a,Davis2014,VanderVelde2012,Khan2012}. \\

\begin{acks}
This study is supported by National Science Foundation (NSF) grant: IIS-1344135 and by National Institute of Health (NIH) grant: LM011972. Contents of this document are the sole responsibility of the authors and do not necessarily represent official views of the NSF/NIH. \\
\end{acks}

\bibliographystyle{ACM-Reference-Format-Journals}
\bibliography{acmsmall-sample-bibfile}
